\documentclass[twocolumn]{aastex63}
\usepackage{lineno}

\usepackage{amsmath}
\usepackage{graphicx}   
\usepackage{bm}         
\usepackage{natbib}
\usepackage{etoolbox}
\usepackage{rotating}
\usepackage{array}
\usepackage{booktabs}
\usepackage{amssymb}

\newcommand{\sm}{$M_\odot$}
\newcommand{\sL}{$L_\odot$}

\newcommand{\co}{C$^{18}$O}

\newcommand{\meta}{CH$_3$OH}

\newcommand{\nhhcho}{NH$_2$CHO}

\newcommand{\kms}{km s$^{-1}$}
\newcommand{\mjybeam}{mJy beam$^{-1}$}
\newcommand{\metd}{CH$_2$DOH}
\newcommand{\medd}{CHD$_2$OH}
\newcommand{\medo}{CH$_3$OD}
\newcommand{\sumire}{$S\mu^2_{\rm \ SUMIRE}$}
\newcommand{\jpl}{$S\mu^2_{\rm \ JPL}$}
\newcommand{\cdms}{$S\mu^2_{\rm \ CDMS}$}

\newcommand{\envelopemetdJPLrange}{[0.41-3.5]$\times$10$^{18}$ cm$^{-2}$}

\newcommand{\centermetdcolumn}{$>$5.2$\times10^{18}$ cm$^{-2}$}
\newcommand{\centermeddcolumn}{$1.1^{+0.4}_{-0.2}\times10^{18}$ cm$^{-2}$}
\newcommand{\centermedocolumn}{$2.0^{+0.5}_{-0.3}\times10^{18}$ cm$^{-2}$}
\newcommand{\envelopemetdrange}{[0.61-4.7]$\times10^{18}$ cm$^{-2}$}
\newcommand{\envelopemeddrange}{[0.11-0.67]$\times10^{18}$ cm$^{-2}$}
\newcommand{\envelopemedorange}{[0.05-0.47]$\times10^{18}$ cm$^{-2}$}

\newcommand{\dhratioJPL}{[0.07-0.26]}

\newcommand{\dohratio}{[0.011-0.06]}
\newcommand{\ddoratioJPL}{[3.5-9.7]}
\newcommand{\dddratioCAL}{[0.20-0.46]}

\newcommand{\dhratio}{[0.09-0.38]}
\newcommand{\ddhratio}{[0.02-0.06]}
\newcommand{\ddoratio}{[4.9-15]}
\newcommand{\dddratio}{[0.14-0.29]}
\newcommand{\dhratiodevide}{[0.03-0.13]}

\newcommand{\chimeta}{0.88}
\newcommand{\massmeta}{0.07}
\newcommand{\rcbmeta}{3}
\newcommand{\incmeta}{70}

\newcommand{\chihcooh}{0.81}
\newcommand{\masshcooh}{0.04}
\newcommand{\rcbhcooh}{7}

\newcommand{\cutwidth}{7}

\newcommand{\metdtemp}{72-188}

\accepted{May 17, 2024}
\submitjournal{ApJ}

\shorttitle{}
\shortauthors{Okoda et al.}

\begin{document}

\title{CH$_3$OH and Its Deuterated Species in the Disk/Envelope System of the Low-Mass Protostellar Source B335}

\correspondingauthor{Yuki Okoda}
\email{yuki.okoda@riken.jp}


\author{Yuki Okoda}
\affiliation{RIKEN Cluster for Pioneering Research, 2-1, Hirosawa, Wako-shi, Saitama 351-0198, Japan}

\author{Yoko Oya}
\affiliation{Center for Gravitational Physics and Quantum Information, Yukawa Institute for Theoretical Physics, Kyoto University, Kyoto, 606-8502, Japan
- Department of Physics, The University of Tokyo, 7-3-1, Hongo, Bunkyo-ku, Tokyo 113-0033, Japan}

\author{Nami Sakai}
\affiliation{RIKEN Cluster for Pioneering Research, 2-1, Hirosawa, Wako-shi, Saitama 351-0198, Japan}

\author{Yoshimasa Watanabe}
\affiliation{Materials Science and Engineering, College of Engineering, Shibaura Institute of Technology, 3-7-5 Toyosu, Koto-ku, Tokyo 135-8548, Japan}

\author{Ana L\'{o}pez-Sepulcre}
\affiliation{Univ. Grenoble Alpes, CNRS, IPAG, 38000 Grenoble, France}
\affiliation{Institut de Radioastronomie Millim\'{e}trique, 38406 Saint-Martin d'H$\grave{e}$res, France}

\author{Takahiro Oyama}
\affiliation{RIKEN Cluster for Pioneering Research, 2-1, Hirosawa, Wako-shi, Saitama 351-0198, Japan}
\author{Shaoshan Zeng}
\affiliation{RIKEN Cluster for Pioneering Research, 2-1, Hirosawa, Wako-shi, Saitama 351-0198, Japan}

\author{Satoshi Yamamoto}
\affiliation{The Graduate University for Advanced Studies SOKENDAI, Shonan Village, Hayama, Kanagawa 240-0193, Japan} 
\affiliation{Research Center for the Early Universe, The University of Tokyo, 7-3-1, Hongo, Bunkyo-ku, Tokyo 113-0033, Japan} 


\begin{abstract}
\par Deuterium fractionation in the closest vicinity of a protostar is important in understanding its potential heritage to a planetary system. Here, we have detected the spectral line emission of \meta\ and its three deuterated species, \metd, \medd, and \medo, toward the low-mass protostellar source B335 at a resolution of 0\farcs03 (5 au) with Atacama Large Millimeter/submillimeter Array.
They have a ring distribution within the radius of 24 au with the intensity depression at the continuum peak.
We derive the column densities and abundance ratios of the above species at 6 positions in the disk/envelope system as well as the continuum peak.
The D/H ratio of \meta\ is $\sim$\dhratiodevide, which is derived by correcting the statistical weight of 3 for \metd.
The $\lbrack$\medd$\rbrack$/$\lbrack$\metd$\rbrack$ ratio is derived to be higher (\dddratio).
On the other hand, the $\lbrack$\metd$\rbrack$/$\lbrack$\medo$\rbrack$ ratio (\ddoratio) is higher than the statistical ratio of 3, and is comparable to those reported for other low-mass sources. 
We study the physical structure on a few au scale in B335 by analyzing the \meta\ (18$_{3,15}-$18$_{2,16}$, A) and HCOOH (12$_{0,12}-$11$_{0,11}$) line emission.
Velocity structures of these lines are reasonably explained as the infalling-rotating motion.
The protostellar mass and the upper limit to centrifugal barrier are thus derived to be 0.03-0.07 \sm\ and $<$7 au, respectively, showing that B335 harbors a young protostar with a tiny disk structure.
Such youth of the protostar may be related to the relatively high $\lbrack$\metd$\rbrack$/$\lbrack$\meta$\rbrack$ ratio.

\end{abstract}


\section{Introduction}\label{intro}
\par The simplest complex organic molecule (COM), methanol (\meta), is widely observed toward protostellar sources to explore their chemical and physical structures.
In star forming clouds, it is thought to be formed through hydrogenation of CO by H atoms on the dust grain surfaces in cold condition during the prestellar phase \citep[e.g.,][]{Watanabe_Kouchi(2002), Geppert et al.(2005), Fuchs et al.(2009)}.
In this phase, desorption of \meta\ into the gas phase is limited to non-thermal processes such as surplus energy release of grain surface reactions. \citep[e.g.,][]{Garrod et al.(2007), Garrod (2013)}.
During the growth of the protostar, \meta\ is released into the gas phase in the inner envelope mainly through the thermal desorption by protostellar heating.
Therefore, the \meta\ emission is used as a tracer of hot core/corino around the protostar \citep[e.g.,][]{Herbst_and_van Dishoeck(2009), Caselli_and_Ceccarelli(2012), Oya et al.(2016), van Gelder et al.(2020),  Imai et al.(2022), Manigand et al.(2020), Ceccarelli et al.(2023), Okoda et al.(2023)}.
The \meta\ emission is also observed in the outflow shocked regions through sputtering and heating processes \citep[e.g.,][]{Bachiller(1997), Codella et al.(2020), Okoda et al.(2021a)}.

Recent high-angular resolution observations reveal the detailed distribution of the \meta\ emission.
For instance, \cite{Oya et al.(2016)} reported that the \meta\ emission in the low-mass protostellar source, IRAS 16293-2422 A, is enhanced in a ring-like structure of the innermost part of the infalling-rotating envelope, which corresponds to the transition zone between the envelope and the disk structure.
Accretion shock and/or the geometrical effect favorable for the protostar heating is proposed as the origin of enhancement \citep{OyaandYamamoto(2020)}.
Accretion shock is also suggested for the enhancement of the \meta\ emission in some sources such as B335 \citep{Okoda et al.(2022)} and [BHB2007] 11 \citep{Vastel et al.(2022)}.
In short, studies on the \meta\ emission provide us with important information connecting the chemical and physical structures of protostellar sources.
\par The deuterated species of \meta\ such as \metd, \medd, CD$_3$OH, and CH$_3$OD have also been detected in various protostellar sources.
Generally, the D/H ratio of molecules (deuterium fractionation) reflects their formation processes as well as physical conditions and evolutionary stages.
In fact, it has extensively been studied with observations of various deuterated species of not only \meta\ but also other molecules over the various regions, such as starless cores \citep[e.g.,][]{Bizzocchi et al.(2014), Ambrose et al.(2021)}, both low-mass and high-mass protostars \citep[e.g.,][]{Bianchi et al.(2017a), Bianchi et al.(2017b), Jorgensen et al.(2018), Taquet et al.(2019), Manigand et al.(2020), van Gelder et al.(2022), Drozdovskaya et al.(2022), Yamato et al.(2023)}, and comets \citep[e.g.,][]{Drozdovskaya et al.(2021), Muller et al.(2022)}.
In these studies, the D/H ratios toward low-mass protostars are found to reach up to 10 \%.
\par In particular, the deuterium fractionation of \meta\ in various protostellar sources would be a key to understand their physical and chemical evolution. Nevertheless, there still remain controversial issues on the \meta\ deuteration. For instance, previous observations suggested a systematic trend of the $\lbrack$\metd$\rbrack$/$\lbrack$\medo$\rbrack$ ratio between low-mass and high-mass protostellar sources.
The ratio in low-mass protostellar sources tends to be higher than the statistical ratio of 3 \citep[e.g.,][]{Bizzocchi et al.(2014), Jorgensen et al.(2018)}, whereas it is much lower than 3 in high-mass protostellar sources \citep[e.g.,][]{Charnley et al.(1997), Belloche et al.(2016), Bogelund et al.(2018), WilkinsandBlake(2022)}.
Thus, more observational efforts for \meta\ and its deuterated spices are awaited to understand the physical meaning of the ratio. 

\par This paper is organized as follows. Our target is introduced in Senction \ref{target_b335}. Some key information on the observation are described in Section \ref{b335_observation}.
In Section \ref{deuterated}, we derive and discuss the column densities and the abundances of deuterated \meta\ in the disk/envelope system.
We show the effect of S$\mu^2$ on the column densities in Section \ref{effect}.
The distributions and velocity maps of the disk/envelope system are shown in Section \ref{maps_velocity_maps}.
Using the infalling-rotating-envelope (IRE) model reported by \cite{Oya et al.(2022)}, we explore kinematics in the disk/envelope system in Section \ref{kinematics}.
In Section \ref{compare}, we discuss the abundance ratios with those in other sources. We summarize the main results in Section \ref{summary}.

\newblock
\section{Target: B335}\label{target_b335}
\par B335 is a Bok globule \citep{Keene et al.(1980)}, with which the Class 0 protostar IRAS 19347+0727 is associated.
The bolometric temperature ($T_{\rm bol}$) is 37 K \citep{Andre et al.(2000)}, while the bolometric luminosity ($L_{\rm bol})$ is 1.6 \sL\ \citep{ Kang et al.(2021)}.
The distance to B335 is reported to be 165 pc, based on the Gaia DR2 parallax data \citep{Watson(2020)}.
In this paper, we employ this distance not only for our results but also for those 
in the previous works.
We adjusted the physical parameter values of the previous works for the new distance (165 pc), since the studies before 2020 employed the distance to B335 of 100 pc, except for \cite{Yen et al.(2015b)} who employed 150 pc.

\par Extensive observations toward this isolated protostellar source have been carried out to develop star-formation studies in terms of both physical and chemical structures \citep[e.g.,][]{Hirano et al.(1988), Evans et al.(2015), Yen et al.(2015b), Bjerkeli et al.(2019), Imai et al.(2016), Imai et al.(2019), Imai et al.(2022), Okoda et al.(2022)}. 
A bipolar outflow extending along the east-to-west direction (P.A. $\sim$90$^\circ$) was found \citep[e.g.,][]{Hirano et al.(1988), Hirano et al.(1992), Stutz et al.(2008), Bjerkeli et al.(2019), Cabedo et al.(2021)}, which is almost in parallel to the plane of the sky. 
Dedicated works have been reported on physical properties of the disk/envelope system and the protostellar mass \citep{Evans et al.(2015), Yen et al.(2015b), Bjerkeli et al.(2019), Imai et al.(2019)}.
The protostellar mass was estimated to be 0.06 \sm\ from imperceptible rotation motion in the \co\ line on the assumption of the infalling-rotating motion by \cite{Yen et al.(2015b)}.
\cite{Imai et al.(2019)} clearly revealed a rotation motion around the protostar in the methanol (\meta) and formic acid (HCOOH) lines with a higher-resolution observation ($\sim$0\farcs1).
They derived the protostellar mass and the radius of the centrifugal barrier to be 0.03$-$0.1 \sm\ and $<$8 au, respectively, assuming the infalling-rotating motion.
Meanwhile, the velocity gradient seen in the \meta\ and SO$_2$ lines are independently reported by \cite{Bjerkeli et al.(2019)}, where the results are consistent with a pure free fall or a Keplerian rotation with the protostellar mass of 0.08 \sm.
They also estimated the disk/envelope mass from the dust continuum emission within 12 au to be 8 $\times$ 10$^{-4}$ \sm.
In B335, a Keplerian disk is still veiled, although recent observations have found it around other young sources even in the Class 0/I stages \citep[e.g.,][]{Ohashi et al.(2014), Aso et al.(2015), Okoda et al.(2018), Oya (2020)}.
\cite{Bjerkeli et al.(2023)} imply that the disk structure just started to form in B335, based on the continuum observation.

The disk/envelope direction is close to the south to north axis. 
We employ the position angle (P.A.) of 5\degr\ for the disk/envelope direction in this paper (Figure \ref{moment}(a)), based on the recent works observing a foot of the outflow with ALMA \citep[P.A. 95\degr:][]{Bjerkeli et al.(2019), Cabedo et al.(2021)}.
\cite{Bjerkeli et al.(2019), Bjerkeli et al.(2023)} suggested the disk/envelope axis to be P.A. 5\degr, and \cite{Cabedo et al.(2021)} reported it to be P.A. 2\degr.
On the other hand, \cite{Oya et al.(2022)} and \cite{Okoda et al.(2022)} employed P.A. 0\degr\ to just take a consistent approach with \cite{Imai et al.(2019)}.
However, only 5 degree difference on the P.A. does not significantly affect any result.

\par Recently, \meta\ and \metd\ were observed in B335 with Atacama Large Millimeter/submillimeter Array (ALMA) at a high resolution of 0.\arcsec03 ($\sim$5 au at $d=$165 pc) by \cite{Okoda et al.(2022)}.
In their study, the temperature structure in the disk/envelope system was the main focus by using \meta, \metd, HCOOH and \nhhcho\ line emission.
The chemical differentiation among molecular distribution were also studied, based on the principal component analysis (PCA).
They found that the \meta\ and deuterated \meta\ species (\metd, \medd, and \medo) have a ring-shaped and extended distribution.
Their velocity structures most likely trace the disk/envelope system, while HCOOH, HNCO, and NH$_2$CHO have a more compact distribution.
Although the column densities of \meta\ and \metd\ were derived simultaneously in derivation of the temperature to show its variation along the major axis of the disk/envelope system, they were not discussed in terms of the D/H ratio. 
Since the intrinsic line intensities, $S\mu^2$ \citep[$S$ is the line strength and $\mu$ the dipole moment responsible for the transition: e.g.,][]{Yamamoto(2017)}, of \metd\ in the database of JPL \citep{Pickett et al.(1998)} was recognized to be inaccurate for their observed transitions in B335, the analysis and discussion of the \meta\ deuteration including \medd\ and \medo\ were left for a separate publication.
\par 
The $S\mu^2$ value is directly related to the individual line intensity and is an important parameter on the line analysis (The details are described elsewhere: \citealp[e.g.,][]{Yamamoto(2017), Oyama et al.(2023)}).
Since the $S\mu^2$ values of \metd\ were recently measured with the laboratory experiment \citep{Oyama et al.(2023)} and those of \medd\ with the theoretical calculation \citep*[][See also \citealt{Drozdovskaya et al.(2022)}]{Coudert et al.(2021)}, reliable derivations of their column densities are now possible by making use of them. 
In this paper, we study the \meta\ deuteration and the velocity structure of the disk/envelope system within a few 10 au scale.



\section{Observation}\label{b335_observation}
\par Single-point ALMA observations toward B335 were carried out with the Band 6 receiver in the four execution blocks of the Cycle 6 operation on June 10, 12, 13, and 23 in 2019. 
The molecular lines analyzed in this paper are summarized in Table \ref{observation}.
The line at 247.2524160 GHz is newly identified in this paper as the \medd\ (4$_{1,2}-$4$_{0,1}$, o$_1$-e$_0$) line, based on the spectroscopic data by \cite{Coudert et al.(2021)} and the database of CDMS \citep{Endres et al.(2016)}.
Since the maximum recoverable scale is 0\farcs3 for these observations, we here focus on the small-scale structure around the protostar.
The synthesized beam size for each line is summarized in Table \ref{observation}. The spectral resolution is 0.544-0.691 \kms, and the root-mean-square noise is 1.0 \mjybeam channel$^{-1}$.
Further observation parameters (calibrators, primary beamwidth, and correlator setups etc.) are described elsewhere \citep{Okoda et al.(2022)}. 
\par The data reduction was performed with Common Astronomy Software Applications package (CASA) 5.8.0 \citep{McMullin et al.(2007)} as well as a modified version of the ALMA calibration pipeline.
We combined four visibility data in the $uv$ plane after phase self-calibration using each continuum data and the application of their solution to the spectral line data.


\section{Deuterated \meta\ species in the Disk/Envelope System}\label{deuterated}
\par We here analyze the \meta, \metd, \medd, and \medo\ lines listed in Table \ref{observation} to study the deuterium fractionation of \meta\ in the disk/envelope system.
Figure \ref{moment0} shows the moment 0 maps of the \meta\ (18$_{3,15}-$18$_{2,16}$, A), \metd\ (4$_{2,2}-$4$_{1,3}$, e$_0$), \medd\ (4$_{1,2}-$4$_{0,1}$, o$_1$-e$_0$), and \medo\ (5$_1-$4$_0$, E) lines with the synthesized beam size (0\farcs030$\times$0\farcs023: Table \ref{observation}).
They have a ring-shaped distribution around the protostar with the intensity depression at the continuum peak due to the high dust opacity.
Similar images are also reported for the \meta, \metd, and \medo\ lines by \cite{Okoda et al.(2022)}, where the beam sizes are smoothed to be 0\farcs034 to perform the PCA.
\par \cite{Okoda et al.(2022)} derived the column densities of \meta\ and \metd\ with each aperture of 0\farcs03 for 9 positions along the midplane of the envelope (P.A. 0\degr) using the $S\mu^2$ values calculated from the line intensities listed in the database of JPL.
A recent experimental measurement for some lines of \metd\ with SUMIRE (Spectrometer Using superconductor MIxer REceiver) by \cite{Oyama et al.(2023)} found that the $S\mu^2$ values in the database of JPL significantly deviate from the measured values.
The experimental $S\mu^2$ values for some lines of \medd\ are now available with SUMIRE very recently (Oyama et al. in prep).
In the following, we first derive the column densities and abundances of \metd\ and \medd\ by using the newly available experimental $S\mu^2$ values (hereafter referred to as $S\mu^2_{\rm \ SUMIRE}$) listed in Table \ref{observation}.
We discuss the dependence of the gas temperature and the column densities on the $S\mu^2$ values later (Section \ref{effect}).

\subsection{Column Densities}\label{column_section}
\par
\begin{figure*}[h!]
\centering
\includegraphics[scale=0.6]{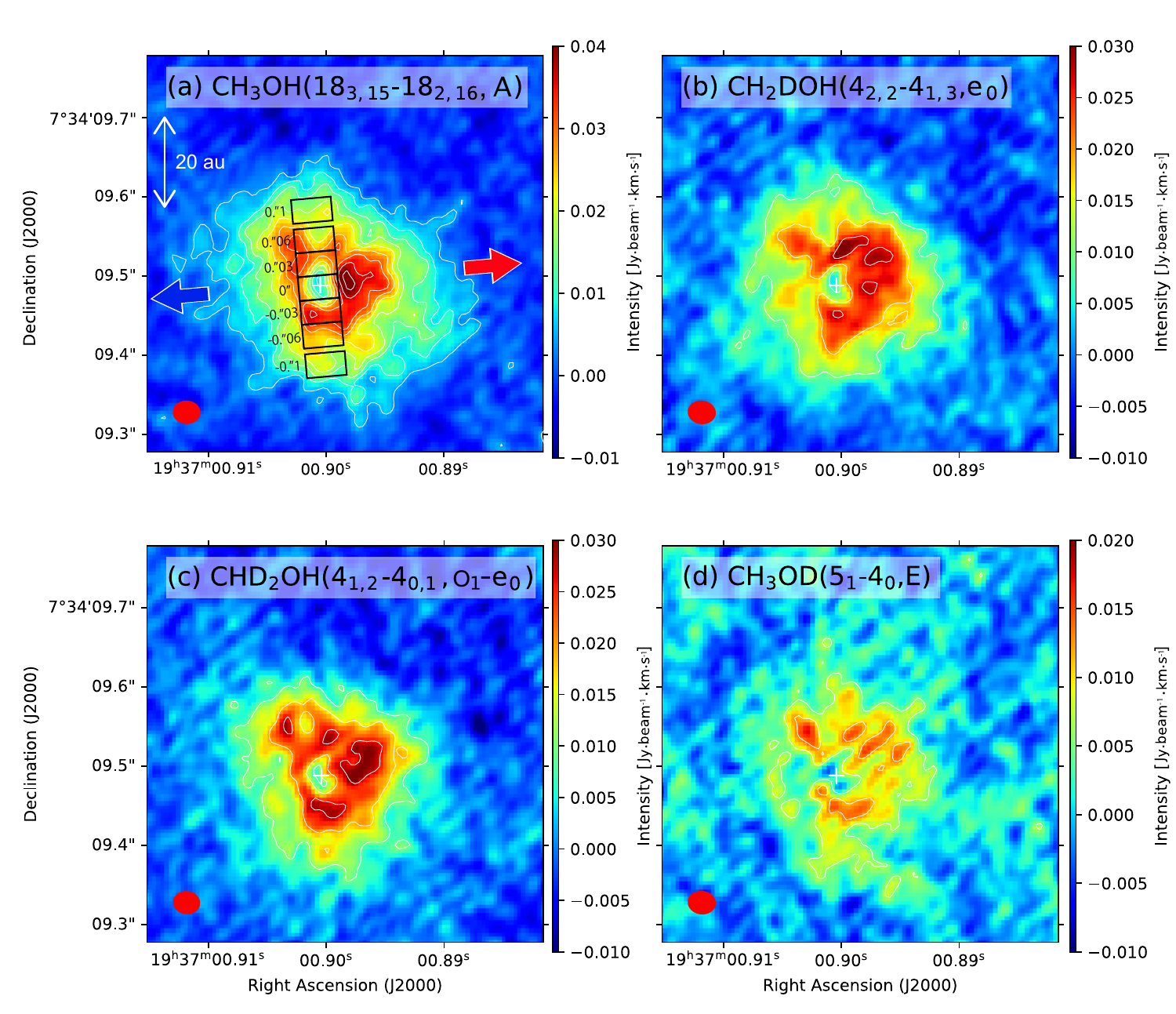}
\caption{Moment 0 maps of the \meta\ (18$_{3,15}-$18$_{2,16}$, A, $E_{\rm u}$=447 K), \metd\ (4$_{2,2}-$4$_{1,3}$, e$_0$, $E_{\rm u}$=38 K), \medd\ (4$_{1,2}-$4$_{0,1}$, o$_1$-e$_0$, $E_{\rm u}$=32 K), and \medo\ (5$_1-$4$_0$, E, $E_{\rm u}$=49 K) emission. The integrated velocity range is from -0.2 \kms\ to 14.5 \kms. The systemic velocity is 8.34 \kms \citep{Yen et al.(2015b)}. Contour levels are every 3$\sigma$ from 3$\sigma$, where $\sigma$ is 1.5 \mjybeam, 3.0 \mjybeam, 3.0 \mjybeam, and 2.0 \mjybeam, respectively.
The spectra are obtained within each rectangle area (0\farcs03$\times$0\farcs05) to calculate the temperature and the column density.
The numbers represent the offsets from the continuum peak (0\farcs0).
The white cross mark show the continuum peak position ($\alpha_{2000}$, $\delta_{2000}$) = (19$^{\rm h}$37$^{\rm m}$00$^{\rm s}$.90$, +7\arcdeg34\arcmin09.$''$49$).
The red ellipse in each map represents the beam size.
The red and blue arrows in the panel (a) show the directions of the red- and blue-shifted outflow.
\label{moment0}}
\end{figure*}

 We derive the column densities of \meta\ and deuterated \meta\ species towards 7 rectangle areas (0\farcs03$\times$0\farcs05) along the disk/envelope direction (P.A. 5\degr) including the continuum peak position, as shown in Figure \ref{moment0}(a).
The rectangle shape is here employed to obtain the spectra with the high signal to noise ratio as much as possible for the disk/envelope system with the apparently thick disk height.
For our derivation, we use the same equations as those reported by \cite{Okoda et al.(2022)}, assuming the local thermodynamic equilibrium (LTE) condition (Appendix \ref{appendix_a}).
In this analysis, we focus on the disk/envelope system within 20 au in radius from the protostar.
The LTE analysis is justified, since the H$_2$ density for that area is roughly estimated to be higher than 10$^8$ cm$^{-3}$, based on its apparent size and the H$_2$ column density of 5.66$\times$10$^{23}$ cm$^{-2}$ reported by \cite{Imai et al.(2016)}.
In addition, the \meta\ emission in our observation selectively traces the warm/hot region because of the maximum recoverable scale of 0\farcs3.
In the derivation, the optical depths of molecular lines and dust emission are considered by assuming that the gas and dust are well mixed and the gas temperature is equal to the dust temperature.
Since the dust continuum emission is intense, we take its effect into the analysis approximately (Appendix \ref{appendix_a}).
\par The best-fit temperatures and column densities are summarized in Table \ref{column}, where the errors are evaluated by the $\chi^2$ analysis.
The offset in Table \ref{column} means the distance from the protostar along the midplane of the disk/envelope system from the southwestern to northeastern direction (P.A. 5\degr), where the offset of 0$\farcs$1 ($\sim$17 au) is almost close to the edge of the distributions of \meta\ and its deuterated species.
Examples of the line parameters of \meta\ at the offset of -0\farcs03 used for the analysis are shown in Appendix \ref{appendix_b}. 
As the upper-state energies of the observed lines of \medd\ are close to one another and there is only one line of \medo\ in our set up, we cannot determine the temperature independently for these two species.
Therefore, in the calculation of the column densities of \medd\ and \medo, we assume the \meta\ temperature obtained for each position, considering the coexisitng nature of \meta\ and its isotopologues (See Figure \ref{moment0}). In addition, the temperature of \meta\ is better determined than that of \metd\ due to the higher signal to noise ratio.
For \medo, we employ the $S\mu^2$ value calculated by \cite{Anderson et al.(1988)}.
\par At the continuum peak, the observed column densities of \meta\ and its isotopologues are seriously affected by the high optical depth of dust continuum emission, as revealed by the central dip of the images in Figure \ref{moment0}.
Only a lower limit can be derived for the column density of \metd\ (\centermetdcolumn), while those of \medd\ and \medo\ are derived to be \centermeddcolumn\ and \centermedocolumn, respectively.
Therefore, we mainly focus on the column densities at the six positions in the envelope.
Among three deuterated species, \metd\ shows the highest column density at all positions, which is in the 
range of \envelopemetdrange.
The column densities of \medd\ and \medo\ are in the range of \envelopemeddrange\ and \envelopemedorange, respectively.
Within the radius of 0\farcs1, the column densities of all species seem to decrease as an increasing offset.
The errors are estimated by using the $\chi^2$ analysis for \meta\ and \metd, where 1$\sigma$ uncertainty is presented as the error for each parameter.
For \medd\ and \medo, we estimate the errors of their column densities, based on the assumed temperature range of $\pm$15 K from the \meta\ temperature.

\subsection{Abundances}\label{abundance_section}
\par
\begin{figure*}[h!]
\centering
\includegraphics[scale=0.6]{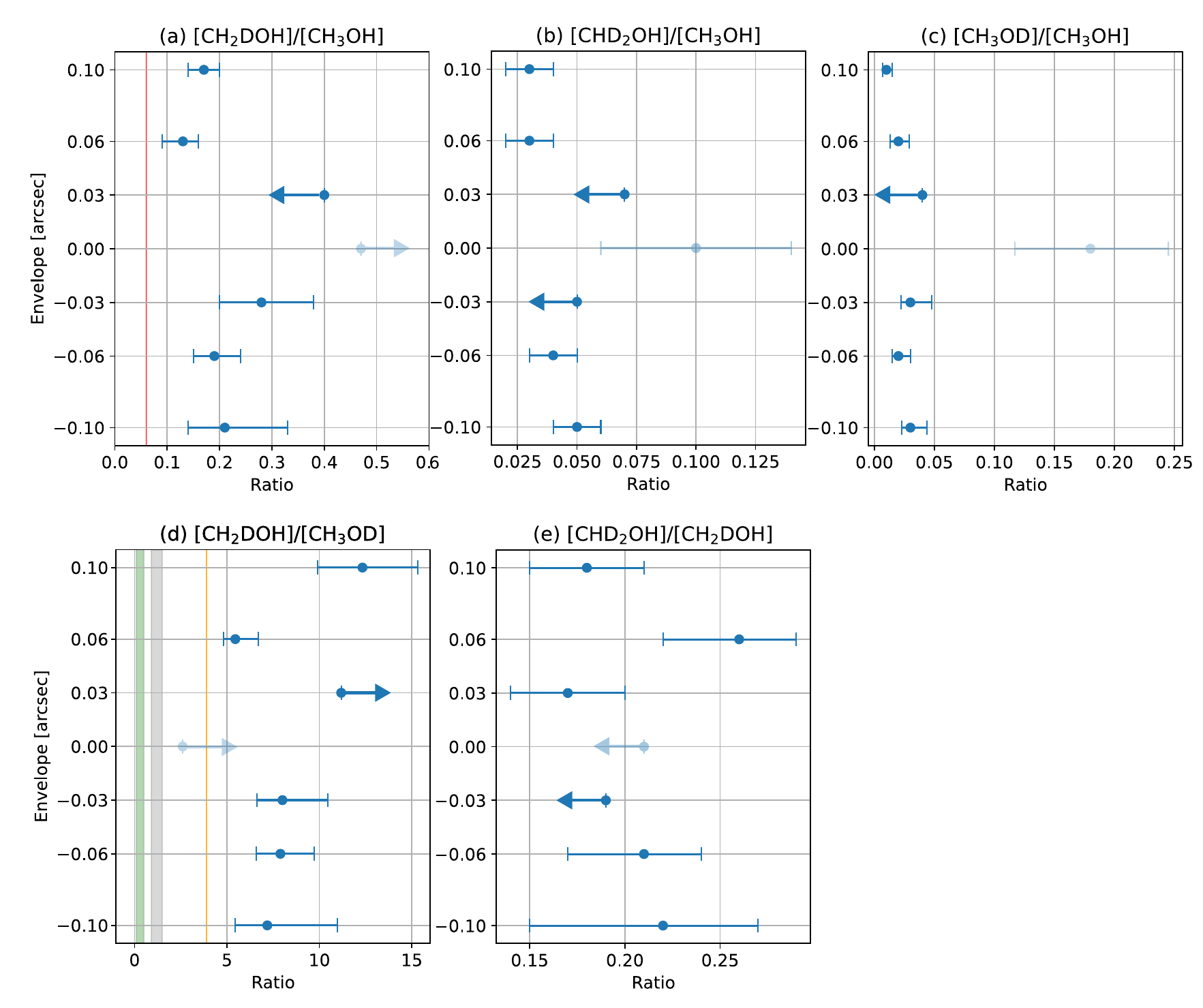}
\caption{(a-e) Abundance ratios along the envelope (P.A. 5$\degr$). Column densities of \metd\ and \medd\ derived from \sumire\ are employed here.
At the continuum peak, the observed column densities are seriously affected by the high optical depth of dust continuum emission, and hence, we focus on the abundances at the six positions in the envelope in this paper.
The left and right arrows show the upper and lower limits of the abundance ratios, respectively.
(a) Red vertical line represents the average in low-mass protostars (0.06) reported by \cite{van Gelder et al.(2022)}.
(d) The ratio in the low-mass protostellar source IRAS 16293-2422 B (3.9) is shown in the orange vertical line for comparison \citep{Jorgensen et al.(2018)}. 
Green and gray areas indicate the ratios in high-mass sources NGC 6334 I (0.1-0.5) \citep{Bogelund et al.(2018)} and Orion KL (0.9-1.5) \citep{Peng et al.(2012), Neill et al.(2013)}, respectively.
\label{d_h}}
\end{figure*}

 Using these derived column densities (Table \ref{column}), we derive the five abundance ratios: $\lbrack$\metd$\rbrack$/$\lbrack$\meta$\rbrack$,
$\lbrack$\medd$\rbrack$/$\lbrack$\meta$\rbrack$,
$\lbrack$\medo$\rbrack$/$\lbrack$\meta$\rbrack$,
$\lbrack$\metd$\rbrack$/$\lbrack$\medo$\rbrack$,
and $\lbrack$\medd$\rbrack$/$\lbrack$\metd$\rbrack$ (Figure \ref{d_h} and Table \ref{abundance}).
As seen in Figure \ref{d_h}, the abundance ratios do not reveal any systematic variation.
Rather, they are roughly constant over the envelope (See also Table \ref{abundance}).
At the continuum peak, all of the ratios have the large uncertainties or just the lower and upper limits due to high dust opacity.
The $\lbrack$\metd$\rbrack$/$\lbrack$\meta$\rbrack$,
$\lbrack$\medd$\rbrack$/$\lbrack$\meta$\rbrack$, and 
$\lbrack$\medo$\rbrack$/$\lbrack$\meta$\rbrack$ ratios in the envelope are in the range of \dhratio, \ddhratio, and \dohratio, respectively (Figures \ref{d_h}(a), (b), and (c)).
In Figures \ref{d_h}(d) and (e), the $\lbrack$\metd$\rbrack$/$\lbrack$\medo$\rbrack$ and $\lbrack$\medd$\rbrack$/$\lbrack$\metd$\rbrack$ ratios are in the range of \ddoratio\ and \dddratio, respectively (See also Table \ref{abundance}).
We discuss the $\lbrack$\metd$\rbrack$/$\lbrack$\meta$\rbrack$, $\lbrack$\medd$\rbrack$/$\lbrack$\metd$\rbrack$, and $\lbrack$\metd$\rbrack$/$\lbrack$\medo$\rbrack$ ratios in Section \ref{compare}.
Although the column densities and the abundances are roughly 
 constant over the disk/envelope system, the temperatures of \meta\ and \metd\ decrease as an increasing distance from the protostar position.
\par The dust optical depths ($\tau_{dust}$)  are summarized in Appendix \ref{appendix_c}.
Note that the $\tau_{dust}$ values obtained in the analyses of the \meta\ and \metd\ lines are slightly different for some positions due to the different temperature.
If we employ the \meta\ temperature for the calculation of \metd, its column density is (0.44-2.8)$\times10^{18}$ cm$^{-2}$ in the disk/envelope system, and the lower limits to the abundance ratios are higher than those when the \metd\ temperature is employed.
In short, the different treatments of the \metd\ temperature do not seriously affect the trend as discussed in Section \ref{compare}.

\section{An Effect of $S\mu^2$ on the Column Densities}\label{effect}
\par In the previous section, we use the experimental value, $S\mu^2_{\rm \ SUMIRE}$, for the derivation of the column density of \metd\ and \medd.
Since the derived column density and rotation temperature depend on the $S\mu^2$ values used, we here note the effects on the results for the different $S\mu^2$ values for the \metd\ (\sumire\ vs. \jpl) and \medd\ (\sumire vs. \cdms) lines.
In Table \ref{observation}, column 4 represents the $S\mu^2$ values calculated from the line intensity in the database of JPL (hereafter referred to as $S\mu^2_{\rm JPL}$) for the \metd\ lines and those taken from the database of CDMS (hereafter referred to as $S\mu^2_{\rm CDMS}$) for the \medd\ lines.
\par The differences between $S\mu^2_{\rm JPL}$ and $S\mu^2$$_{\rm SUMIRE}$ are conspicuous for \metd:
$S\mu^2$$_{\rm SUMIRE}$ is 0.6-0.8 times smaller than $S\mu^2$$_{\rm JPL}$.
The temperature and the column density toward the offset of -0\farcs03 from the continuum peak are derived with $S\mu^2$$_{\rm JPL}$ to be 162$^{+12}_{-8}$ K and 2.5$^{+0.8}_{-0.4}\times$10$^{18}$ cm$^{-2}$, respectively.
When we use $S\mu^2$$_{\rm SUMIRE}$, they are derived to be 161$^{+8}_{-6}$ K and 3.6$^{+1.1}_{-0.6}\times$10$^{18}$ cm$^{-2}$, respectively.
The former column density is about 1.4 times smaller than the latter column density, while the temperatures are almost the same as each other.
At the other positions, the column densities of \metd\ derived with $S\mu^2_{\rm JPL}$ is also 1.4-1.6 times smaller than those derived with \sumire, as shown in Table \ref{column}, resulting in the range of the column densities and the ratios as: $N$(\metd)=\envelopemetdJPLrange, $\lbrack$\metd$\rbrack$/$\lbrack$\meta$\rbrack$=\dhratioJPL, and $\lbrack$\metd$\rbrack$/$\lbrack$\medo$\rbrack$=\ddoratioJPL.
Note that some of the \metd\ column densities are different from those reported by \cite{Okoda et al.(2022)} beyond the errors and the effect of the $S\mu^2$ difference, particularly at the continuum peak position.
This is because the circle area with the diameter of 0\farcs03 was employed for their derivation.
\par On the other hand, a difference between $S\mu^2$$_{\rm CDMS}$ and $S\mu^2$$_{\rm SUMIRE}$ for the \medd\ lines is relatively small (Table \ref{observation}).
The $S\mu^2$$_{\rm SUMIRE}$ values are about 1.2 times larger than the $S\mu^2$$_{\rm CDMS}$ values at most for the observed lines.
In this case, the upper limits to the column densities at the offset of -0\farcs03 are similar to each other: they are $<$0.77$\times$10$^{18}$ cm$^{-2}$ and $<$0.68$\times$10$^{18}$ cm$^{-2}$ with the $S\mu^2$$_{\rm CDMS}$ and $S\mu^2$$_{\rm SUMIRE}$ values, respectively, where we assume the \meta\ temperature of 186$\pm$15 K.
Therefore, a choice of the $S\mu^2$$_{\rm CDMS}$ and $S\mu^2$$_{\rm SUMIRE}$ values does not cause serious differences on the abundance ratios in our observation (Table \ref{abundance}).
The $\lbrack$\medd$\rbrack$/$\lbrack$\metd$\rbrack$ ratio is mostly affected by the difference on the $S\mu^2$ value of \metd.
Using the \jpl\ value for \metd\ and the \cdms\ value for \medd, we obtain \dddratioCAL\ (Table \ref{abundance}).

\par It should be noted that an effect of $S\mu^2$ values on the column densities depend on which lines are used for the derivation.
In other words, different lines have different differences between the S$\mu^2_{\rm JPL}$ and $S\mu^2$$_{\rm SUMIRE}$ values or between the S$\mu^2_{\rm CDMS}$ and $S\mu^2$$_{\rm SUMIRE}$ values.
Therefore, an effect of the $S\mu^2$ values on the column densities in any other observations would be different from those reported here.
Thus, we need a special care for a comparison of column densities and their ratios with those in other sources reported previously.

\section{Distributions and Velocity Maps of the disk/envelope system}\label{maps_velocity_maps}
\par It is of fundamental importance for characterization of this source to verify the protostellar mass and the presence/absence of the disk structure, which may relate to the evolutionary stage of the protostar.
As mentioned in Section \ref{target_b335}, various efforts toward this direction have been reported.
In particular, \cite{Imai et al.(2019)} studied velocity structures of the disk/envelope system with the \meta\ (12$_{6,7}-$13$_{5,8}$, E, $E_{\rm u}=$360 K) and HCOOH (12$_{0,12}-$11$_{0,11}$, $E_{\rm u}=$83 K) lines at a resolution of 0\farcs1.
Since the resolution of our data is higher by a factor of 3-4, we re-investigate the kinematic structure of the disk/envelope system with these molecular species.
\par
\begin{figure}[h!]
\centering
\includegraphics[scale=1]{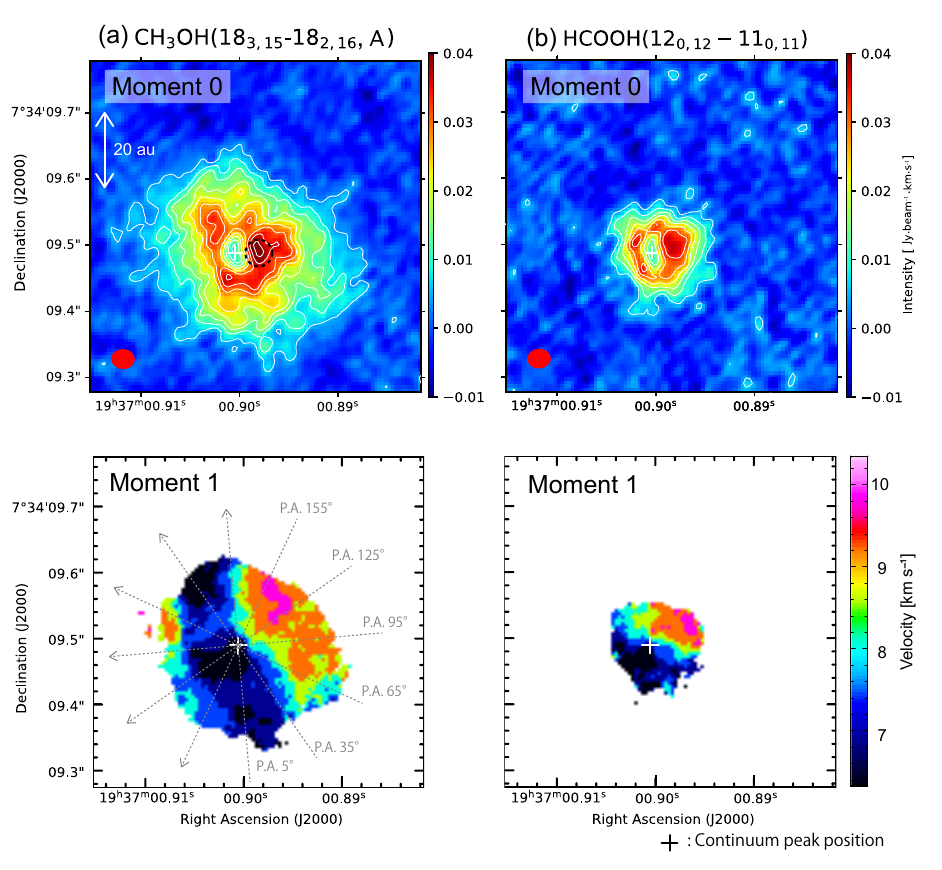}
\caption{
(a,b) Moment 0 and 1 maps of the \meta\ (18$_{3,15}-$18$_{2,16}$, A) and HCOOH (12$_{0,12}-$11$_{0,11}$) emission in the upper and lower panels, respectively. The moment 0 map of \meta\ is the same as Figure \ref{moment0}(a), where the black dotted circle indicates the area covered by the SiO emission with previous work, as mentioned in Section \ref{maps_velocity_maps}. Contour levels for HCOOH are every 3$\sigma$ from 3$\sigma$, where $\sigma$ is 1.8 \mjybeam. For the moment 0 and 1 maps, the integrated velocity range are from -0.2 \kms\ to 14.5 \kms\ and from 1.6 \kms to 13.3 \kms, respectively. The systemic velocity is 8.34 \kms \citep{Yen et al.(2015b)}. The gray dotted arrows in the moment 1 map of \meta\ show the directions of the PV diagrams of Figures \ref{pvmeta}, \ref{pvhcoohire}, and \ref{pvhcoohkep}. The white cross marks represent the continuum peak ($\alpha_{2000}$, $\delta_{2000}$) = (19$^{\rm h}$37$^{\rm m}$00$^{\rm s}$.90$, +7\arcdeg34\arcmin09.$''$49$). The red ellipses represent the beam size.
\label{moment}}
\end{figure}

 The upper panels of Figure \ref{moment} show the moment 0 maps of \meta\ (18$_{3,15}-$18$_{2,16}$, A, $E_{\rm u}=$447 K) and HCOOH (12$_{0,12}-$11$_{0,11}$), respectively.
The velocity range for the integration is from -0.2 \kms\ to 14.5 \kms, where the systemic velocity is 8.34 \kms \citep{Yen et al.(2015b)}.
The moment 0 map of \meta\ is the same as Figure \ref{moment0}(a).
We choose this line of \meta\ for the kinematic analysis in the next section because of the relatively high signal to noise ratio.
In addition, we can avoid the contamination from the outflow as much as possible with the line, since it has the relatively high upper-state energy among the \meta\ lines in our observation.
In the upper panel of Figure \ref{moment}(b), the HCOOH line shows a more compact distribution, which would selectively trace the velocity structure of the inner disk/envelope system.
To study the inner structure, we also explore the velocity structure of the HCOOH emission in this paper.
\par The lower panels of Figure \ref{moment} show the moment 1 maps of \meta\ and HCOOH, respectively, where the velocity range is from 1.6 \kms\ to 13.3 \kms.
The velocity gradient can be seen in the both of the moment 1 maps.
The velocity gradient in the map of \meta\ is almost along the west to east axis, whereas that of HCOOH is rather close to the south to north axis.
The outflow and disk/envelope directions are suggested to be nearly along the west to east (P.A. 95\degr) and south to north axes (P.A. 5\degr), respectively \citep[e.g,][]{Bjerkeli et al.(2023)}.
Hence, the observed velocity structure has a gradient along the different direction from the suggested disk/envelope system, particularly for the \meta\ line.
As \cite{Imai et al.(2019)} and \cite{Oya et al.(2022)} pointed out, this can occur for the contribution of the infalling-rotating motion, which will be discussed later (Section \ref{meta_kine}).

\par The moment 0 maps of \meta\ and HCOOH show a ring-shaped distribution in spite of the almost edge-on configuration suggested by \cite{Bjerkeli et al.(2019)} and \cite{Evans et al.(2023)}. 
The \meta\ distribution is more extended in height of the disk/envelope system than the HCOOH distribution as well as in radius.
Part of the \meta\ emission can be caused by the interaction with the outflow as shown in the lower panel of Figure \ref{moment}(b) in addition to the infalling-motion effect above and below the midplane of the disk/envelope system.
Such a stratified molecular distribution was also reported in the other low-mass source: HH212 by \cite{Lee et al.(2022)}.
The HCOOH emission would be less affected by the outflow interaction than the \meta\ emission and provide us with a better estimate of the disk height.
Therefore, the apparent scale height of the disk/envelope system is roughly be estimated to be about 20 au based on the HCOOH maps for the analysis in the next section (Figure \ref{moment}(b)).
\par Moreover, in B335, the SiO emission was reported to have an extended distribution along the east to west axis near the protostar ($<$ 0\farcs1) by \cite{Bjerkeli et al.(2019)}. SiO is well known as a shock tracer \citep[e.g.,][]{Mikami et al.(1992), Bachiller(1997)}, and could trace the launching point of the outflow or the accretion shock of infalling material \citep{Imai et al.(2019)}. 
Its distribution covers the area where both of the \meta\ and HCOOH line intensities are enhanced around the west of the protostar, as shown in the dotted black circle of Figure \ref{moment}(a).
In terms of the detection of SiO, the intensity enhancement would be caused by a shock on the surface area of the disk/envelope system rather than the protostellar heating.

\section{Kinematics of the Disk/Envelope System with FERIA}\label{kinematics}
\par In this section, we analyze the \meta\ and HCOOH data to determine the protostellar mass ($M_{\rm star}$), the centrifugal barrier ($r_{\rm CB}$), and the inclination ($i$) of the disk/envelope system, the inner radius ($R_{\rm in}$) with the aid of a general-purpose computer code FERIA (Flat Envelope model with Rotation and Infall under Angular momentum conservation) developed by \cite{Oya et al.(2022)}.
The physical meaning of $R_{\rm in}$ depends on the model, and hence, we will specify it for each case later.
We perform the $\chi^2$ test to obtain the best-fit model parameters on the molecular-line data cube, where the intensities are higher than 3$\sigma$ noise level; 1.5 \mjybeam\ and 1.8 \mjybeam\ for \meta\ and HCOOH.
For the models, we set a height of the disk/envelope system to be 20 au based on the apparent height in the map of HCOOH, as mentioned in Section \ref{maps_velocity_maps}.

\subsection{\meta\ (18$_{3,15}-$18$_{2,16}$, A)}\label{meta_kine}
\par Based on the distributions in the moment 0 and 1 maps (Figure \ref{moment}(a)) and the analyses presented by \cite{Imai et al.(2019)}, the \meta\ emission is expected to trace an infalling-rotating envelope (IRE).
To explore the velocity structure of the \meta\ (18$_{3,15}-$18$_{2,16}$, A) emission, we prepare the IRE models by taking $M_{\rm star}$, $r_{\rm CB}$, and $i$ as the free parameters.
$R_{\rm in}$ here represents the inner radius of IRE ($R_{\rm in}$) traced by the \meta\ emission, which is assumed to be equal to $r_{\rm CB}$.
$R_{\rm out}$ indicates the outer radius of IRE ($R_{\rm out}$) traced by the \meta\ emission, which is fixed to be 24 au according to the observed emission.
The parameter ranges are summarized in Table \ref{para}.

\par The disk/envelope system was suggested to be the almost edge-on configuration (0\degr\ for face-on: \citealp[e.g.,][]{Bjerkeli et al.(2019), Imai et al.(2019), Evans et al.(2023)}), and hence, we set the range from 70\degr\ to 90\degr\ for $i$.
The position angles (P.A.) of the disk/envelope is set to be 5\degr\ \citep{Bjerkeli et al.(2019), Bjerkeli et al.(2023), Cabedo et al.(2021)}. 
We compare the model with the observation on the cube data to calculate the reduced $\chi^2$ value.
Then, we obtain the best-fit parameters at the reduced $\chi^2$ values of \chimeta, where $M_{\rm star}$, $r_{\rm CB}$($=R_{\rm in}$), and $i$ are \massmeta \sm, \rcbmeta\ au, and \incmeta\degr, respectively.
\par \cite{Imai et al.(2019)} reported $M_{\rm star}$ of 0.03-0.1 \sm\ and $r_{\rm CB}<$8 au.
With the same data, \cite{Oya et al.(2022)} obtained the best-fit parameters of $M_{\rm star}$=0.03 \sm\ and $r_{\rm CB}=$2 au through the $\chi^2$ tests (the reduced $\chi^2$ value$=$0.77) for the infalling-rotating model.
They also analyzed the observed data with a Keplerian model, resulting in the best-fit parameters of $M_{\rm star}$=0.07 \sm\ and $R_{\rm in}=$2 au at a reduced $\chi^2$ value of 0.78.
\cite{Bjerkeli et al.(2019)} independently derived $M_{\rm star}$ to be 0.08 \sm\ from the analysis of the PV diagrams of \meta\ and SO$_2$ at a similar resolution to our observation, assuming a Keplerian rotation.
Thus, the small $M_{\rm star}$ and $r_{\rm CB}$ values obtained in this study are consistent with those reported in the previous works.
\par
\begin{figure}[h!]
\centering
\includegraphics[scale=0.8]{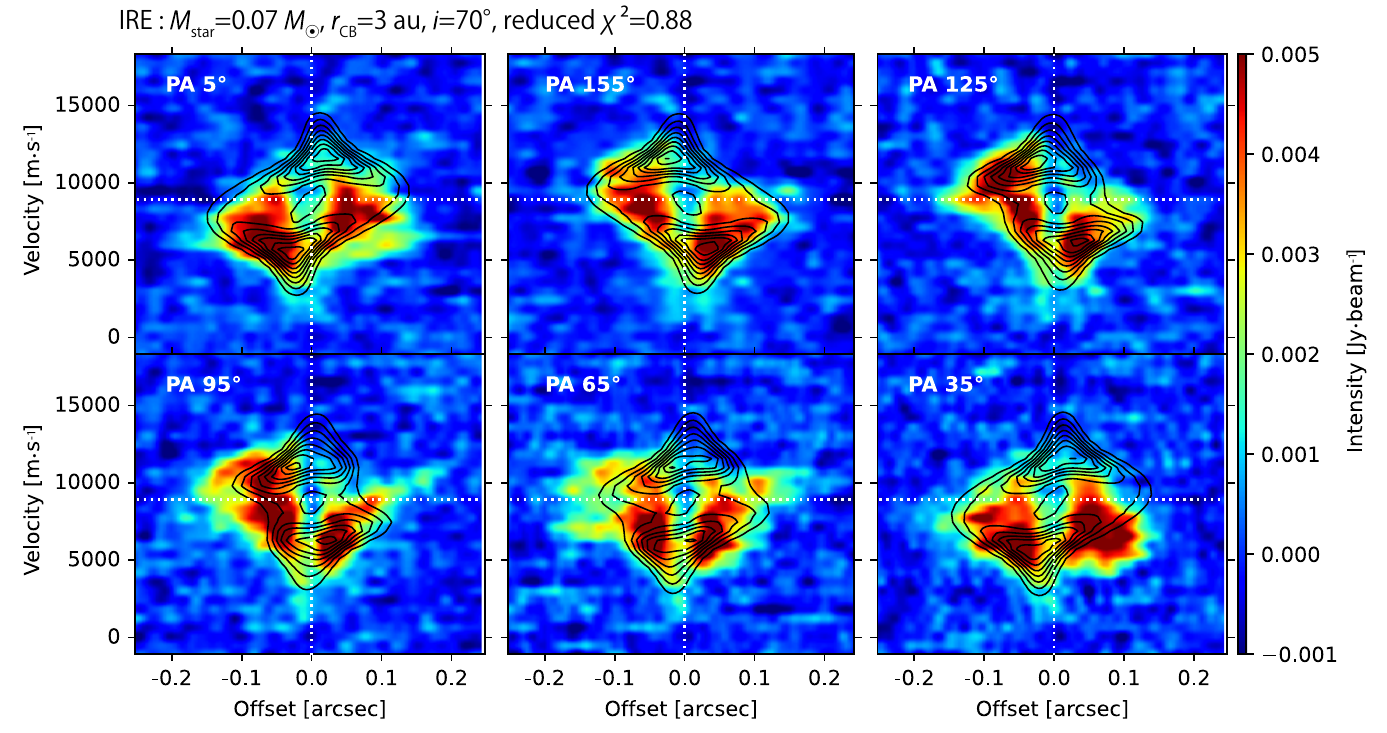}
\caption{PV diagrams of \meta\ (18$_{3,15}-$18$_{2,16}$, A) in colors and the best-fit model obtained by the $\chi^2$ analysis for the infalling-rotating envelope (IRE) model with FERIA in contours. The best-fit parameters for the IRE model are the protostellar mass of 0.07\sm, the outer radius of 24 au, the inner radius and the centrifugal barrier of 3 au, and the inclination of 70\degr\ (90\degr\ for edge-on), where the reduced $\chi^2$ value is 0.88.
Dotted horizontal lines show the systemic velocity of 8.34 \kms \citep{Yen et al.(2015b)}, and dotted vertical lines show the protostar position.}
\label{pvmeta}
\end{figure}

 Figure \ref{pvmeta} shows the PV diagrams of the best-fit model in contours compared with those of \meta\ in colors, where the cutting width of \cutwidth\ au for PVs.
Directions of the PV diagrams are shown as the arrows in Figure \ref{moment}(a).
The model can almost reproduce the major features of the observation, although some emission is veiled by the dust continuum at the continuum peak (offset of 0\farcs0).
In addition, the redshifted component is weaker than the blueshifted one in Figure \ref{pvmeta}. This may be due to the self-absorption in the infalling gas (i.e., inverse P-Cygni profile).
An increase of the velocity around the offsets of 0\farcs1-0\farcs2, as seen in the panels of P.A. 95\degr\ and 65\degr, is likely a contribution of an outflow interaction.

\subsection{HCOOH (12$_{0,12}-$11$_{0,11}$)}
\par We obtain $r_{\rm CB}$ of \rcbmeta\ au in the analysis for the \meta\ line.
This means that if there were a rotationally supported disk inside the infalling-rotating envelope traced by \meta, its radius would be \rcbmeta\ au.
This size cannot fully be resolved by the resolution of our data ($\sim$5 au).
Since the HCOOH emission has a distribution of 10 au in radius that is larger than the obtained $r_{\rm CB}$, it seems to overlap with part of the \meta\ distribution.
As well, the moment 1 map (Figure \ref{moment}(b)) looks similar to the infalling-rotating motion \citep{Oya et al.(2022)}.
Nevertheless, we examine the following two cases to explore its velocity structure by using the model on the cube data.
One is the case that HCOOH traces an infalling-rotating motion with $R_{\rm out}$ of 10 au in the inner part of the disk/envelope system.
In this case, there could be a very small rotationally supported disk within $R_{\rm in}$ ($r_{\rm CB}$).
Another case is that a Keplerian disk with the radius of 10 au (=$R_{\rm out}$=$r_{\rm CB}$) is traced by HCOOH, where the gas infalling on the above and below disk mid-plane toward the protostar is traced by \meta.
The two cases are schematically shown in Figures \ref{pvcase}. 
In the following cases of Sections \ref{irecase} and \ref{kepcase}, $R_{\rm out}$ and $R_{\rm in}$ refer to the outer and inner radii of the structure traced by the HCOOH emission, respectively.

\subsubsection{Case 1: Infalling-rotating envelope (IRE)}\label{irecase}
\begin{figure}[h!]
\centering
\includegraphics[scale=0.8]{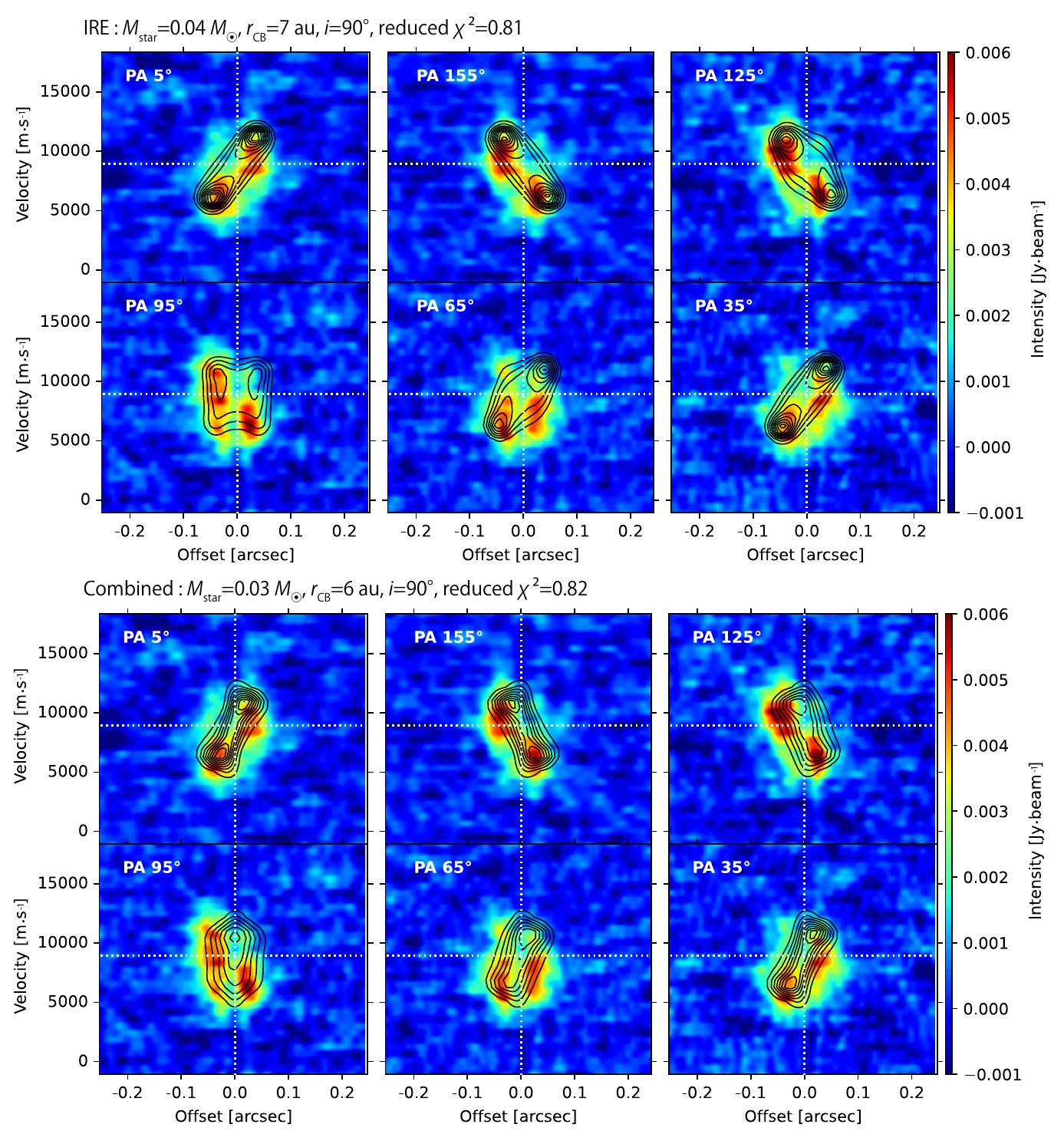}
\caption{PV diagrams of HCOOH (12$_{0,12}-$11$_{0,11}$) in colors and the best-fit model obtained by the $\chi^2$ analysis for the infalling-rotating envelope (IRE) model (upper) and the combined model (lower) with FERIA in contours. The parameters for the IRE model are the protostellar mass of 0.04\sm, the outer radius of 10 au, the inner radius and the centrifugal barrier of 7 au, and the inclination of 90\degr\ (90\degr\ for edge-on), where the reduced $\chi^2$ value is 0.81. Those for the combined model are the protostellar mass of 0.03\sm, the outer radius of 10 au, the inner radius and the centrifugal barrier of 6 au, and the inclination of 90\degr\ (90\degr\ for edge-on), where the reduced $\chi^2$ value is 0.82.
Dotted horizontal lines show the systemic velocity of 8.34 \kms \citep{Yen et al.(2015b)}, and dotted vertical lines show the protostar position.}
\label{pvhcoohire}
\end{figure}

We here consider the IRE model with $R_{\rm out}$ of 10 au as well as the combined model.
In the combined model, the IRE model and the Keplerian rotation model are considered simultaneously, where the Keplerian disk is assumed to exist inward of the centrifugal barrier of the IRE (i.e., for $r_{\rm Kepler}$ $<$ $r_{\rm CB}$).
At first, we prepare the IRE models using $M_{\rm star}$, $r_{\rm CB}$($=R_{\rm in}$), and $i$ summarized in Table \ref{para}, and perform a reduced $\chi^2$ test for HCOOH on the cube data.
The best-fit value of $M_{\rm star}$ is \masshcooh\ \sm, where the reduced $\chi^2$ value is \chihcooh.
The other best-fit parameters are: $r_{\rm CB}$($=R_{\rm in}$)=\rcbhcooh\ au, and $i=$90\degr.
These parameters are slightly different from those in the \meta\ analysis (Section \ref{meta_kine}).
$r_{\rm CB}$ of \rcbhcooh\ au is within the range reported by \cite{Imai et al.(2019)}.
The larger value of $r_{\rm CB}$ than that obtained by the \meta\ analysis ($r_{\rm CB}=$3 au) may be caused by the weaker intensities near the continuum peak.
In Figure \ref{pvhcoohire}, PV diagrams of the best-fit models are overlaid on those of the HCOOH emission.
Most of the observed features are reproduced by the IRE model.
If we set $r_{\rm CB}$($=R_{\rm in}$) to be \rcbmeta\ au obtained in the \meta\ analysis, the best-fit parameters of $M_{\rm star}$ and $i$ are 0.03 \sm\ and 90\degr, respectively, where the reduced $\chi^2$ value is 0.84.
\par For the combined models, we fix the inner radius ($R_{\rm in}$) of the Keplerian part to be 1 au to reduce the number of the free parameters.
The outer radius ($R_{\rm out}$) and the ranges for the other parameters ($M_{\rm star}$, $r_{\rm CB}$, and $i$) are the same as those of the IRE model.
The $\chi^2$ analysis on the data cube yields the best-fit parameters as: $M_{\rm star}=$0.03 \sm, $r_{\rm CB}(=R_{\rm in})=$6 au, and $i=$90\degr, where the reduced $\chi^2$ value is 0.82.
Then, we cannot confirm the disk structure in this observation, because of the insufficient resolution and the relatively weak emission of the central part due to the dust opacity effect.

\subsubsection{Case 2: Keplerian rotation}\label{kepcase}
\begin{figure}[h!]
\centering
\includegraphics[scale=1.4]{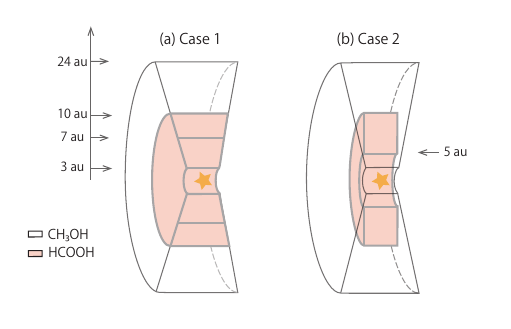}
\caption{Schematic illustration of the disk/envelope system of B335. The HCOOH and \meta\ emitting region areas are enclosed by colors and by solid and dashed lines, respectively. (a) The inner envelope traced by the HCOOH emission and the overall envelope traced by the \meta\ emission. (b) Stratified molecular distribution is considered. A possible Keplerian disk traced by the HCOOH emission and the infalling gas envelope traced by the \meta\ emission motion.\label{pvcase}}
\end{figure}

The stratified molecular distributions mentioned above being considered, the disk structure could be embedded in the infalling gas traced by the \meta\ emission.
An infalling-rotating motion on the surface of the disk/envelope system near the protostar would be traced by \meta\ (Figure \ref{pvcase}(b)).
In this case, HCOOH might trace a Keplerian disk with the radius of 10 au.
With this situation in mind, we prepare the models with a Keplerian rotion in the FERIA code.
The size of the disk, $R_{\rm out}$, is fixed to be 10 au, and the free parameters are $M_{\rm star}$, $R_{\rm in}$, and $i$, which are varied in the ranges of 0.02-0.08 \sm, 1-6 au, and 70-90\degr, respectively, as summarized in Table \ref{para}.
We obtain the best-fit parameters of $M_{\rm star}=$0.05 \sm, $R_{\rm in}=$5 au, and $i=$90\degr\ with the reduced $\chi^2$ value of 0.82.
$i$ suggests a completely edge-on disk, while $M_{\rm star}$ is similar to that in the \meta\ analysis based on the IRE model. 
The reduced $\chi^2$ value in the Case 2 is similar to that in Case 1.
\par In short, our observation is not able to determine which of Case 1 and Case 2 is more appropriate for the HCOOH distribution.
Nevertheless, we can conclude that B335 has a very small protostellar mass and a disk structure smaller than the HCOOH emission.
Our analysis on the HCOOH emission suggests that we need a higher-angular resolution and/or other molecular lines specifically tracing a more compact distribution to find a Keplerian disk.

\section{Deuteration and Physical Environment}\label{compare}
\par Based on our analysis in Section \ref{kinematics}, the \meta\ emission traces an infalling-rotating motion within 24 au in radius.
The free-fall time is very short, which is roughly estimated to be only $\sim$ 10 yr, where the infall velocity of 5 \kms\ is assumed at the radius of 10 au.
Even if the dynamical time scale is estimated by the Kepler time, it is about 100 yr.
As a result, there would not be enough time to change the isotopic ratios in the inner envelope by gas-phase chemical reactions.
Furthermore, our analysis with FERIA implies a tiny disk structure with the radius of $<$7 au ($r_{\rm CB}$).
Considering these results, we here discuss the abundance ratios $\lbrack$\metd$\rbrack$/$\lbrack$\meta$\rbrack$, $\lbrack$\medd$\rbrack$/$\lbrack$\metd$\rbrack$, and $\lbrack$\metd$\rbrack$/$\lbrack$\medo$\rbrack$ in B335 and compare them with those in other sources reported previously.

\subsection{$\lbrack$\metd$\rbrack$/$\lbrack$\meta$\rbrack$}\label{sec:d_h_ratio}
\par The $\lbrack$\metd$\rbrack$/$\lbrack$\meta$\rbrack$ ratio can be converted to the D/H ratio by dividing by 3, because there are three equivalent H atoms of the methyl group for the D substitution.
The D/H ratio was reported to be up to 0.1 toward low-mass protostars \citep[e.g.,][]{Taquet et al.(2019), van Gelder et al.(2020), van Gelder et al.(2022)}.
 \cite{van Gelder et al.(2022)} discussed the D/H ratio of $\lbrack$\metd$\rbrack$/$\lbrack$\meta$\rbrack$ among large samples including low-mass and intermediate sources as well as high-mass sources, where the lines of \meta\ isotopologues, $^{13}$\meta\ and CH$_3$$^{18}$OH, as well as normal \meta\ lines are used.
Since most of the samples of low-mass protostars in their paper have the ratio much lower than 0.1, its average in low-mass protostars was reported to be $\lbrack$\metd$\rbrack$/$\lbrack$\meta$\rbrack$$\sim$0.06 (D/H$\sim$0.02).
 It is higher than the ratios in high-mass protostars.
The $\lbrack$\metd$\rbrack$/$\lbrack$\meta$\rbrack$ ratio in B335 is \dhratio\ (D/H$\sim$\dhratiodevide), which tends to be relatively high even among other low-mass protostars and low-mass prestellar cores (D/H$\sim$0.034$\pm$0.019) \citep{van Gelder et al.(2022)}.
This result may be related to the young evolutionary stage of the B335 protostar.
Since not enough time would have passed to decrease the D/H ratio to its equilibrium ratio at the current temperature in B335, the D/H ratio in the prestellar stage may be converted to some extent.
Nevertheless, the derived D/H ratio could be due to underestimating the column density of \meta.
The optical depths of the \meta\ lines are derived to be relatively high, as summarized in Appendix \ref{appendix_d}. The D/H ratio of \meta\ could thus be consistent with those of the previous reports. Anyway, further comfirmation  of the D/H ratio by optically thin isotopologues is awaited: a few $^{13}$\meta\ lines are in the observed spectral windows, but none of them provide further constraints on the column density.
\par Note that the \jpl\ values are used for \metd\ in the previous studies.
Even if we employed the \jpl\ values in our calculation, the D/H ratio in B335 would be smaller only by a factor of $\sim$1.5.
Therefore, the above trend on $\lbrack$\metd$\rbrack$/$\lbrack$\meta$\rbrack$ does not change even for this case (Table \ref{abundance}).

\subsection{$\lbrack$\medd$\rbrack$/$\lbrack$\metd$\rbrack$}\label{medd_metd}
The $\lbrack$\medd$\rbrack$/$\lbrack$\metd$\rbrack$ ratio (D/H ratio) of \dddratio\ in B335 is comparable to those in the low-mass protostellar sources IRAS 16293$-$2422 Source A (0.20$\pm$0.07) and B (0.25$\pm$0.09) \citep{Drozdovskaya et al.(2022)}.
For the low-mass protostellar sources, B1-c and Serpens S68N, \cite{van Gelder et al.(2022)} derived the ratios 0.13$\pm$0.02 and 0.12$\pm$0.05, respectively, on the assumption of the temperature of 150 K. They are slightly lower than or comparable to our result.
On the other hand, our result in B335 is lower than the ratios in the other low-mass protostellar sources, NGC1333 IRAS 2A (0.70$\pm$0.26), IRAS 4A (0.56$\pm$0.22) \citep{Taquet et al.(2019)}, and HOPS373SW (0.75) \citep{Lee et al.(2023)}.
Recently, the \medd\ and \metd\ lines were detected in the prestellar sources as well, the ratios being 0.8$\pm$0.4 in H-MM1 and 0.5$\pm$0.3 in L694-2 \citep{Lin et al.(2023)}.
Furthermore, the D/H ratio from \medd\ and \metd\ is reported to be higher than that from \metd\ and \meta\ in above protostellar and prestellar sources (Appendix \ref{comparison_app}).
Our results in B335 follow this trend even in the close vicinity of protostar ($<$24 au).
Note that the upper range of the D/H ratio of \meta\ is comparable to the lower range of the $\lbrack$\medd$\rbrack$/$\lbrack$\metd$\rbrack$ ratio. There may be the effect of the underestimated \meta\ column density due to the high optical depth in our observation.
Since the actual \meta\ column density could become higher, the $\lbrack$\metd$\rbrack$/$\lbrack$\meta$\rbrack$ ratio would be lower, resulting in the values in the range of D/H ratio from \medd\ and \metd\ higher than that from \metd\ and \meta.

\subsection{$\lbrack$\metd$\rbrack$/$\lbrack$\medo$\rbrack$}
\par The $\lbrack$\metd$\rbrack$/$\lbrack$\medo$\rbrack$ ratio in B335 is found to be as high as \ddoratio\ (Table \ref{abundance}).
The similar values were found in other low-mass sources \citep[e.g.,][]{Parise et al.(2004), Parise et al.(2006),Bizzocchi et al.(2014),  Jorgensen et al.(2018)}.
In the low-mass Class 0 protostellar source, IRAS 16293-2422 Source B, the ratio is derived to be 3.9 at the position shifted from the continuum peak as one beam to avoid the dust opacity effects with ALMA \citep{Jorgensen et al.(2018)}.
\cite{Bizzocchi et al.(2014)} reported its lower limit of 10 in the external layers of the prestellar core L1544.
On the other hand, the ratios in some high-mass star forming regions are lower than 1.0 \citep[e.g.,][]{Jacq et al.(1993), Ratajczak et al.(2011), Belloche et al.(2016), Bogelund et al.(2018), WilkinsandBlake(2022)}.
Thus, it is recognized that there is a systematic trend between low-mass and high-mass protostars.
The physical environment would thus be one of the important effects on the $\lbrack$\metd$\rbrack$/$\lbrack$\medo$\rbrack$ ratio.

\par Even if the \sumire\ values for \metd\ were used for the calculation in the previously reported $\lbrack$\metd$\rbrack$/$\lbrack$\medo$\rbrack$ ratios, the trend would still be significant for most cases.
For the strong lines of \metd, except for some c-type transitions or extremely high excitation ones, the differences of $S\mu^2$ between the JPL database and the experiment are within a factor of a few. 
It is therefore hard to explain the reported difference of the $\lbrack$\metd$\rbrack$/$\lbrack$\medo$\rbrack$ ratios more than an order of magnitude only by the different $S\mu^2$ values used in each study.  
\par For \meta, the methyl group has three equivalent H atoms, while the hydroxy group only one H atom. 
If the hydrogenation of CO randomly occurs without isotope effect, the $\lbrack$\metd$\rbrack$/$\lbrack$\medo$\rbrack$ ratio should be 3.
This is so-called the statistical ratio, which is predicted by conventional gas-grain chemical models \citep[e.g.,][] {Charnley et al.(1997),Osamura et al.(2004)}.
However, it is known that the deuterium fractionation of the CH$_3$ group can be enhanced through abstraction of H by the D atom and subsequent addition of D on the grain surface \citep{Hidaka et al.(2009)}. 
This additional process occurs efficiently if the atomic D/H ratio on dust grains are high in cold conditions. This mechanism would make the $\lbrack$\metd$\rbrack$/$\lbrack$\medo$\rbrack$ ratio higher than the statistical ratio of 3. 
Such a high abundance ratio was also suggested with chemical models by \cite{Kulterer et al.(2022)}.
According to their studies, \medo\ is formed at a warm-up stage, although \metd\ could be inherited from prestellar stage to protostellar stage.

\par The chemical pathway for \medo\ is still controversial, where many studies have been done by the chemical models \citep[e.g.,][]{Charnley et al.(1997), Rodgers(2002), Osamura et al.(2004), Taquet et al.(2014), WilkinsandBlake(2022), Kulterer et al.(2022)} and the laboratory experiments \citep[e.g.,][]{Nagaoka et al.(2005), Ratajczak et al.(2009)}.
\cite{Bogelund et al.(2018)} and \cite{Taquet et al.(2019)} compared the $\lbrack$\metd$\rbrack$/$\lbrack$\medo$\rbrack$ ratios in high-mass source (NGC 6334 I: 0.1-0.5) and low-mass sources (IRAS 2A: 0.8-3.6 and IRAS 4A: 1.2-5.3) with the chemical models presented by \cite{Taquet et al.(2012), Taquet et al.(2013), Taquet et al.(2014)}.
The models follow deuteration of \meta\ in cold dense cores (10-40 K), resulting in the ratio close to or a little higher than 3.
Hence, the observed trend for the high-mass protostellar source, NGC6334I, is not reproduced.
On the other hand, \cite{Faure et al.(2015a)} reported a model that can reproduce the observational trend of the high and low $\lbrack$\metd$\rbrack$/$\lbrack$\medo$\rbrack$ ratios in low-mass and high-mass protostellar sources, respectively, considering the pathway exchanging between water and \meta\ in icy mantles of dust grains.
\cite{Kulterer et al.(2022)} also pointed out the same mechanism.
\par Recently, \cite{WilkinsandBlake(2022)} discussed rapid D-H exchange in methanol-containing ices depending on the temperature, based on its chemical model and the observation toward the high-mass star forming region Orion KL.
According to their observation, the column density of \medo\ starts to rise steeply at $\sim$110 K and keeps increasing until before $\sim$185 K.
This result is consistent with their chemical model representing the rapid variation of the gas-phase \medo\ column density due to the D-H exchange between water and \meta\ on ices.
In fact, the similar increase of the \medo\ column density as increasing temperature up to around 200 K is also supported experimentally \citep{Souda et al.(2003), Kawanowa et al.(2004)}.
\cite{WilkinsandBlake(2022)} suggested that the low $\lbrack$\metd$\rbrack$/$\lbrack$\medo$\rbrack$ ratio in Orion KL is caused by the increase of the \medo\ column density at the temperature ($\sim$100-200 K), while understandings of the D-H exchange to produce \medo\ at a temperature higher than $\sim$ 200 K is controversial.
\par If the high temperature were the only key factor for increasing \medo\ (i.e, low $\lbrack$\metd$\rbrack$/$\lbrack$\medo$\rbrack$ ratio), the $\lbrack$\metd$\rbrack$/$\lbrack$\medo$\rbrack$ ratio in the inner envelope of B335 should be as low as in high-mass star forming regions because of the high temperature condition (107-217 K for \meta).
However, this simple thought contradicts our observational result. 
Therefore, the ratio may also depend on the physical environment at the prestellar core phase.
The time scale that the D-H exchange reaches steady state in Orion KL is suggested to be $<$10$^3$ yr \citep{Faure et al.(2015a)}.
The free-fall and dynamical time scales in B335 on a few 10 au scale seems to be shorter than the chemical reaction timescale.
Alternatively, there might be an effect by high cosmic rays in the envelope of B335, where the cosmic-ray ionization rate is reported to be 10$^{-16}$-10$^{-14}$ s$^{-1}$ by \cite{Cabedo et al.(2021)}.
This rate is higher than usually assumed in chemical models \citep[10$^{-17}$ s$^{-1}$:][]{Yamamoto(2017)}.
Apparently, we need more observations of other protostellar sources in various physical environments to clarify the link between $\lbrack$\metd$\rbrack$/$\lbrack$\medo$\rbrack$ and physical environment.

\section{Summary}\label{summary}
\par We study the \meta\ and its deuterated species in terms of the abundance ratios and  kinematics of the disk/envelope system in the low-mass protostellar source B335 at a high resolution (0$\farcs$03$\sim$5 au) with ALMA.
Main results are summarized below.\\
\\
1. We analyze 4 lines of \metd, 3 lines of \medd, and 1 line of \medo\ as well as 6 lines of \meta\ in Band 6 originally observed by \cite{Okoda et al.(2022)}.
The \medd\ (4$_{1,2}-$4$_{0,1}$, o$_1$-e$_0$) line emission is newly identified in this paper, which was previously unidentified by \cite{Okoda et al.(2022)}.
These molecular lines are ring-shaped and extended within 24 au in radius around the protostar.\\
\\
2. The five abundance ratios ($\lbrack$\metd$\rbrack$/$\lbrack$\meta$\rbrack$, $\lbrack$\medd$\rbrack$/$\lbrack$\meta$\rbrack$,
$\lbrack$\medo$\rbrack$/$\lbrack$\meta$\rbrack$,\\
$\lbrack$\metd$\rbrack$/$\lbrack$\medo$\rbrack$,
and $\lbrack$\medd$\rbrack$/$\lbrack$\metd$\rbrack$) are almost constant over the envelope. 
This likely originates from the short dynamical timescale (10-100 yr) in the inner envelope compared to the chemical timescale ($\sim10^{4}$ yr).\\
\\
3. We examine the effect of $S\mu^2$ values on the calculation for column densities and temperatures of \metd\ and \medd, because the calculated $S\mu^2$ values (\jpl or $S\mu^2_{\rm CDMS}$) and the experimental $S\mu^2$ values by \cite{Oyama et al.(2023)} (\sumire) are different.
Since \sumire\ for the \metd\ lines that we use is 0.6-0.8 times smaller than \jpl, its column density derived with \sumire\ is about 1.5 times larger than that with \jpl.
On the other hand, the difference is smaller for \medd, which does not make a large effect on our column density derivations.\\
\\
4. With the aid of the infalling-rotating-envelope model FERIA, we find that the \meta\ (18$_{3,15}-$18$_{2,16}$, A) emission traces an infalling-rotating envelope with $M_{\rm star}$, $r_{\rm CB}$($=R_{\rm in}$), and $i$ are \massmeta \sm, \rcbmeta\ au, and \incmeta\degr, respectively. 
Additionally, we study the HCOOH (12$_{0,12}$$-$11$_{0,11}$) line with the IRE model and the Keplerian model to examine the existence of the disk structure.
The disk structure is not definitively found, and these analyses just suggest a tiny disk structure smaller than 7 au in radius.
The low $M_{\rm star}$ and the very small disk structure imply a very young protostellar stage of this source.\\
\\
5. The $\lbrack$\metd$\rbrack$/$\lbrack$\meta$\rbrack$ ratio in B335 (\dhratio) is relatively high among other low-mass protostellar (D/H$\sim$0.02) and prestellar sources (D/H$\sim$0.034$\pm$0.019) \citep{van Gelder et al.(2022)}.
\metd\ produced on dust grains at the prestellar phase seems to remain in the gas phase after desorption.
This result may be related to the youth of the B335 protostar, where a fresh gas is being supplied from the outer cold envelope. 
Further confirmation using optically thin isotopologue lines is necessary.\\
\\
6. In B335, the $\lbrack$\medd$\rbrack$/$\lbrack$\metd$\rbrack$ ratio (\dddratio) is higher than the D/H ratio of \meta\  ($\lbrack$\metd$\rbrack$/$\lbrack$\meta$\rbrack$/3=[0.03-0.13]).
This feature follows the trend among the other low-mass protostars.
\\
\\
7. The $\lbrack$\metd$\rbrack$/$\lbrack$\medo$\rbrack$ ratio in B335 (\ddoratio) is higher than the statistical weight of 3. This result further supports the systematic trend between low-mass and high-mass sources in previous studies.
Even under the high temperature condition (\meta: 107-217K and \metd: \metdtemp\ K) in B335, the $\lbrack$\metd$\rbrack$/$\lbrack$\medo$\rbrack$ ratio is still high. This fact means that the ratio would not only depend on the temperature but on current and past physical environments. 

\acknowledgments
This paper makes use of the following ALMA data set:
ADS/JAO.ALMA\# 2018.1.01311.S (PI: Muneaki Imai). ALMA is a partnership of the ESO (representing its member states), the NSF (USA) and NINS (Japan), together with the NRC (Canada) and the NSC and ASIAA (Taiwan), in cooperation with the Republic of Chile.
The Joint ALMA Observatory is operated by the ESO, the AUI/NRAO, and the NAOJ. The authors thank to the ALMA staff for their excellent support.
This project is supported by a Grant-in-Aid from Japan Society for the Promotion of Science (KAKENHI: No. 19H05069, 19K14753, and 22K20390.
Y. Okoda thanks RIKEN Special Postdoctoral Researcher Program (Fellowships) for financial support.


{}

\begin{table}[htbp!]
\centering
\caption{\centering{Analyzed Molecular Lines$^a$} \label{observation}}
\scalebox{0.8}{
\begin{tabular}{ccccccccc}
\hline \hline
Molecule&Transition & Frequency & $S\mu^2$$^b$ &$S\mu^2$$_{\rm SUMIRE}$$^c$ &log$_{10}$ $A$$^d$ &log$_{10}$ $A_{\rm SUMIRE}$$^e$ & $E_{\rm u}$$k^{-1}$ & Synthesized beam size\\
&&(GHz)&($D^2$)&($D^2$)&&&(\rm K)&\\
\hline  
\meta  & 18$_{3,15}-$18$_{2,16}$, A  & 247.610918 & 17.358 & - & -4.081 & -  & 446.6  & 0$\farcs$033$\times$0$\farcs$027 (P.A. -2.6°) \\
\meta  & 21$_{3,18}-$21$_{2,19}$, A  & 245.223019 & 20.623 & - & -4.084 & -  & 585.8  & 0$\farcs$033$\times$0$\farcs$028 (P.A. -6.9°) \\
\meta  & 12$_{6,7}-$13$_{5,8}$, E & 261.704409 & 2.131 & - & -4.75 & -  & 359.8  & 0$\farcs$032$\times$0$\farcs$026 (P.A. 4.6°) \\
\meta  & 17$_{3,14}-$17$_{2,15}$, A  & 248.282424 & 16.315 & - & -4.081 & -  & 404.8  & 0$\farcs$032$\times$0$\farcs$026 (P.A. -9.4°) \\
\meta  & 4$_{2,2}-$5$_{1,5}$, A  & 247.228587 & 1.086 & - & -4.673 & -   & 60.9  & 0$\farcs$033$\times$0$\farcs$027 (P.A. -5.1°) \\
\meta  & 2$_{1,1}-$1$_{0,1}$, E  & 261.805675 & 1.334 & - & -4.254 & -   & 28.0  & 0$\farcs$032$\times$0$\farcs$026 (P.A. 4.3°) \\
\metd  & 10$_{2,8}-$10$_{1,9}$, o$_1$ & 244.9888456 & 3.439 & 2.22 & -4.552 & -4.74  & 153.3  & 0$\farcs$033$\times$0$\farcs$028 (P.A. -7.4°) \\
\metd  & 4$_{2,2}-$4$_{1,3}$, e$_0$  & 244.8411349 & 2.540 & 1.80 & -4.317 & -4.47 & 37.6  & 0$\farcs$033$\times$0$\farcs$028 (P.A. -7.5°) \\
\metd  & 5$_{2,4}-$5$_{1,5}$, e$_0$  & 261.6873662 & 4.006 & 3.01 & -4.119 & -4.24 & 48.3  & 0$\farcs$032$\times$0$\farcs$026 (P.A. 4.6°) \\
\metd  & 3$_{2,1}-$3$_{1,2}$, e$_0$ & 247.6257463 & 2.360 & 1.68 & -4.225 & -4.37  & 29.0  & 0$\farcs$033$\times$0$\farcs$027 (P.A. -2.6°) \\
\medd$^f$  & 6$_{1,1}-$5$_{1,1}$, o$_1$  & 246.1432950 & 4.821 & 5.22 & -4.191 & -4.16  & 52.9  & 0$\farcs$033$\times$0$\farcs$027 (P.A. -4.4°) \\
\medd$^f$  & 6$_{1,2}-$5$_{1,2}$, e$_0$ & 246.2530390 & 4.759 & 5.36 & -4.196 & -4.14 &44.6  & 0$\farcs$033$\times$0$\farcs$027 (P.A. -4.4°) \\
\medd$^f$  & 4$_{1,2}-$4$_{0,1}$, o$_1$-e$_0$ & 247.2524160 & 5.655 & 6.86 & -3.956 & -3.87 & 31.8  & 0$\farcs$032$\times$0$\farcs$027 (P.A. -7.7°) \\
\medo  & 5$_1-$4$_0$, E  & 245.142988 & 3.8 & - & -4.8 & -  & 37.3  & 0$\farcs$033$\times$0$\farcs$028 (P.A. -6.9°) \\
HCOOH &12$_{0,12}-$11$_{0,11}$&262.1034810&24.157&- &  -3.694 &-&82.8&0$\farcs$032$\times$0$\farcs$026 (P.A. 4.4$^{\circ}$)\\
\hline
\end{tabular}}
\begin{flushleft}
\tablecomments{
$^a$ Line parameters are taken from CDMS \citep{Endres et al.(2016)} and JPL \citep{Pickett et al.(1998)} except for CH$_3$OD.
The parameters for \medo\ are taken from \cite{Anderson et al.(1988)} and \cite{Duan et al.(2003)}.\\
$^b$ Taken from CDMS for \meta\ and \medd\ and calculated from the intensity listed in JPL for \metd. For \medo, we employ the data reported by \cite{Anderson et al.(1988)}. 
$^c$ Taken from \cite{Oyama et al.(2023)} for \metd. Provided by T. Oyama for \medd. These values are the experimental data. The uncertainty is 10$\%$ for the \sumire\ value.
$^d$ Einstein coefficients calculated from \jpl\ or \cdms.
$^e$ Einstein coefficients calculated from \sumire.
$^f$ The quantum numbers follow the definition of CDMS.}
\end{flushleft}
\end{table}

\begin{table*}[ht]
\centering
\caption{Temperatures and Column Densities in the Disk/envelope System} \label{column}
\scalebox{0.8}{
\begin{tabular}{cccccc}
\hline \hline
& &\multicolumn{2}{c}{$S\mu^2_{\rm JPL}$ or $S\mu^2_{\rm CDMS}$$^e$} & \multicolumn{2}{c}{\sumire}\\
\hline
Molecules &Offsets$^a$ &    $T$ (K)  & $N$ (10$^{18}$cm$^{-2}$) &$T$ (K) &  $N$ (10$^{18}$cm$^{-2}$)  \\
\hline 
\meta$^b$  & 0.1 & 107$^{+2}_{-2}$ & 4.4$^{+0.5}_{-0.5}$ &-&-\\
&0.06 & 142$^{+3}_{-4}$ & 9.0$^{+3.0}_{-1.7}$ &-&-\\
&0.03 & 164$^{+6}_{-6}$ & $>$9.6 &-&-\\
&0$^{d}$ & 217$^{+2}_{-2}$ & 11$^{+5}_{-2}$ &-&-\\
&-0.03 & 186$^{+2}_{-2}$ & 13$^{+4}_{-3}$ &-&-\\
&-0.06 & 146$^{+4}_{-3}$ & 8.0$^{+1.7}_{-1.3}$&-&- \\
&-0.1 & 124$^{+5}_{-5}$ & 3.7$^{+1.1}_{-0.8}$&-&- \\
\hline 
\metd$^b$  &  0.1 & 72$^{+3}_{-1}$ & 0.47$^{+0.06}_{-0.06}$ & 72$^{+2}_{-2}$ & 0.74$^{+0.10}_{-0.10}$ \\
           & 0.06 & 119$^{+12}_{-8}$ & 0.85$^{+0.09}_{-0.06}$ & 118$^{+8}_{-8}$ & 1.2$^{+0.2}_{-0.1}$ \\
           & 0.03 & 143$^{+3}_{-3}$ & 2.6$^{+0.9}_{-0.4}$ & 143$^{+3}_{-3}$ & 3.8$^{+0.9}_{-0.6}$ \\
           & 0$^{d}$ & 188$^{+8}_{-6}$ & $>$3.5 & 188$^{+8}_{-4}$ & $>$5.2 \\
           & -0.03 & 162$^{+12}_{-8}$ & 2.5$^{+0.8}_{-0.4}$ & 161$^{+8}_{-6}$ & 3.6$^{+1.1}_{-0.6}$ \\
            &-0.06 & 135$^{+30}_{-16}$ & 1.00$^{+0.20}_{-0.11}$ & 132$^{+22}_{-12}$ & 1.5$^{+0.3}_{-0.2}$ \\
           & -0.1 & 106$^{+36}_{-14}$ & 0.52$^{+0.26}_{-0.10}$ & 104$^{+26}_{-12}$ & 0.79$^{+0.41}_{-0.18}$ \\
            
\hline 
\medd$^c$ & 0.1 & 107$^{+15}_{-15}$  & 0.15 $^{+ 0.02 }_{ -0.02 }$  & 107$^{+15}_{-15}$ & 0.13 $^{+ 0.02 }_{ -0.02 }$ \\
&0.06 & 142$^{+15}_{-15}$ & 0.35 $^{+ 0.04 }_{ -0.03 }$ & 142$^{+15}_{-15}$ & 0.31 $^{+ 0.03 }_{ -0.03 }$   \\
&0.03 & 164$^{+15}_{-15}$  & 0.71 $^{+ 0.05 }_{ -0.01 }$& 164$^{+15}_{-15}$ & 0.629 $^{+ 0.042 }_{ -0.002 }$   \\
&0$^{d}$ & 217$^{+15}_{-15}$    & 1.2 $^{+ 0.6 }_{ -0.2 }$& 217$^{+15}_{-15}$ & 1.1 $^{+ 0.4 }_{ -0.2 }$ \\
&-0.03 & 186$^{+15}_{-15}$   & $<$0.77& 186$^{+15}_{-15}$  & $<$0.68 \\
&-0.06 & 146$^{+15}_{-15}$   & 0.35 $^{+ 0.03 }_{ -0.02 }$& 146$^{+15}_{-15}$& 0.31 $^{+ 0.03 }_{ -0.02 }$   \\
&-0.1 & 124$^{+15}_{-15}$    & 0.20 $^{+ 0.02 }_{ -0.01 }$& 124$^{+15}_{-15}$& 0.17 $^{+ 0.02 }_{ -0.01 }$  \\
\hline 
\medo$^c$ &0.1 & 107$^{+15}_{-15}$  & 0.06 $^{+ 0.01 }_{ -0.01 }$ &-&-\\
&0.06 & 142$^{+15}_{-15}$  & 0.22 $^{+ 0.02 }_{ -0.03 }$ &-&-\\
&0.03 & 164$^{+15}_{-15}$  & $<$0.34 &-&-\\
&0$^{d}$ & 217$^{+15}_{-15}$  & 2.0 $^{+ 0.5 }_{ -0.3}$&-&- \\
&-0.03 & 186$^{+15}_{-15}$  & 0.45 $^{+ 0.02 }_{ -0.01 }$ &-&-\\
&-0.06 & 146$^{+15}_{-15}$  & 0.19 $^{+ 0.02 }_{ -0.02 }$&-&- \\
&-0.1 & 124$^{+15}_{-15}$  & 0.11 $^{+ 0.01 }_{ -0.01 }$ &-&-\\
\hline
\end{tabular}}
\begin{flushleft}
\tablecomments{$^a$ Offsets from the continuum peak position along the envelope (P.A. 5$\degr$). 
$^b$ The errors are estimated by using the $\chi^2$ analysis.
$^{c}$ These temperatures are assumed as that of \meta. For the error estimation of \medd\ and \medo, we assume the temperature error of $\pm$15 K. These values are derived from the spectra for the rectangle area (0$\farcs$03$\times$0$\farcs$05) shown in Figure \ref{moment0}(a).
$^{d}$ At the continuum peak, the observed column densities are seriously affected by the high optical depth of dust continuum emission, and hence, we focus on the values at the six positions in the envelope in this paper.
$^{e}$ For \metd\ and \medd, the column densities are calculated with the \jpl\ and \cdms\ values, respectively.}
\end{flushleft}
\end{table*}

\begin{table}[ht]
\centering
\caption{Abundance Ratios in the Disk/envelope System} \label{abundance}
\scalebox{0.7}{
\begin{tabular}{ccccccc}
\hline
\hline  
$S\mu^2_{\rm JPL}$ or $S\mu^2_{\rm CDMS}$$^a$ &Offset$^b$  & $\lbrack$\metd$\rbrack$/$\lbrack$\meta$\rbrack$ &$\lbrack$\medd$\rbrack$/$\lbrack$\meta$\rbrack$  & $\lbrack$\medo$\rbrack$/$\lbrack$\meta$\rbrack$&$\lbrack$\metd$\rbrack$/$\lbrack$\medo$\rbrack$ & $\lbrack$\medd$\rbrack$/$\lbrack$\metd$\rbrack$   \\
 \hline
&0.1 & 0.11$^{+0.02}_{-0.02}$ & 0.03$^{+0.01}_{-0.01}$ & 0.014$^{+0.005}_{-0.003}$ & 7.8$^{+1.9}_{-1.5}$ & 0.32$^{+0.06}_{-0.06}$ \\
&0.06 & 0.09$^{+0.02}_{-0.02}$ & 0.04$^{+0.01}_{-0.01}$ & 0.024$^{+0.008}_{-0.007}$ & 3.9$^{+0.7}_{-0.4}$ & 0.41$^{+0.05}_{-0.05}$ \\
&0.03 & $<$0.27 & $<$0.07 & $<$0.04 & $>$7.7 & 0.27$^{+0.04}_{-0.07}$ \\
&0$^{e}$ & $>$0.32 & 0.11$^{+0.06}_{-0.04}$ & 0.18$^{+0.06}_{-0.06}$ & $>$1.8 & $<$0.34 \\
&-0.03 & 0.19$^{+0.07}_{-0.05}$ & $<$0.06 & 0.04$^{+0.02}_{-0.01}$ & 5.6$^{+1.8}_{-0.9}$ & $<$0.31 \\
&-0.06 & 0.12$^{+0.03}_{-0.03}$ & 0.04$^{+0.01}_{-0.01}$ & 0.024$^{+0.008}_{-0.005}$ & 5.3$^{+1.2}_{-0.8}$ & 0.35$^{+0.05}_{-0.06}$ \\
&-0.1 & 0.14$^{+0.07}_{-0.04}$ & 0.05$^{+0.01}_{-0.01}$ & 0.03$^{+0.01}_{-0.01}$
& 4.7$^{+2.4}_{-1.0}$ & 0.38$^{+0.07}_{-0.13}$ \\
\hline
{\sumire}$^c$ &Offset$^{b,d}$& $\lbrack$\metd$\rbrack$/$\lbrack$\meta$\rbrack$ &$\lbrack$\medd$\rbrack$/$\lbrack$\meta$\rbrack$  & $\lbrack$\medo$\rbrack$/$\lbrack$\meta$\rbrack$&$\lbrack$\metd$\rbrack$/$\lbrack$\medo$\rbrack$ & $\lbrack$\medd$\rbrack$/$\lbrack$\metd$\rbrack$   \\
 \hline
&0.1 & 0.17$^{+0.03}_{-0.03}$ & 0.03$^{+0.01}_{-0.01}$ & - & 12$^{+3}_{-2}$ & 0.18$^{+0.03}_{-0.03}$ \\
&0.06 & 0.13$^{+0.03}_{-0.04}$ & 0.03$^{+0.01}_{-0.01}$ & - & 5.5$^{+1.3}_{-0.6}$ & 0.26$^{+0.03}_{-0.04}$ \\
&0.03 & $<$0.40 & $<$0.07 & - & $>$11 & 0.17$^{+0.03}_{-0.03}$ \\
&0$^{e}$ & $>$0.47 & 0.10$^{+0.04}_{-0.04}$ & - & $>$2.6 & $<$0.21 \\
&-0.03 & 0.28$^{+0.1}_{-0.08}$ & $<$0.05 & - & 8.0$^{+2.5}_{-1.4}$ & $<$0.19 \\
&-0.06 & 0.19$^{+0.05}_{-0.04}$ & 0.04$^{+0.01}_{-0.01}$ & - & 7.9$^{+1.8}_{-1.3}$ & 0.21$^{+0.03}_{-0.04}$ \\
&-0.1 & 0.21$^{+0.12}_{-0.07}$ & 0.05$^{+0.01}_{-0.01}$ & - & 7.2$^{+3.8}_{-1.7}$ & 0.22$^{+0.05}_{-0.07}$ \\
\hline
\end{tabular}}
\begin{flushleft}
\tablecomments{
$^a$ For \metd\ and \medd, the abundances are derived by using the column densities calculated with the \jpl\ and \cdms\ values, respectively.
$^b$ Offsets from the continuum peak position along the envelope (P.A. 5$\degr$). 
$^c$ For \metd\ and \medd, the abundances are derived by using the column densities calculated with the \sumire\ value.
$^d$The abundance ratios are plotted in Figure \ref{d_h}.
$^{e}$ At the continuum peak, the observed column densities are seriously affected by the high optical depth of dust continuum emission, and hence, we focus on the abundances at the six positions in the envelope in this paper.}

\end{flushleft}
\end{table}


\begin{table}[ht]
\centering
\caption{Free Parameters in the Reduced $\chi^2$ Test on the Cube Data} \label{para}
\scalebox{1}{
\begin{tabular}{cccccccc}
\hline \hline
 && $M_{star}$ & $r_{\rm CB}$& $i$&$R_{\rm in}$&$R_{\rm out}$& Reduced\\
Model&Molecule&(\sm) & (au)&(\degr) & (au)& (au)&$\chi^2$ Values\\
\hline
 IRE&\meta\ & 0.02-0.08  & 2-9& 70-90& $r_{\rm CB}$&24\\
(Best-fit)&&0.07 & 3 & 70&&&0.88\\
 IRE&HCOOH & 0.02-0.08  & 2-9& 70-90& $r_{\rm CB}$&10&\\
(Best-fit)&&0.04& 7& 90& & & 0.81\\
Combined&HCOOH & 0.02-0.08  & 2-9& 70-90& 1&10&\\
(Best-fit)&& 0.03& 6 & 90& & & 0.82\\
Kepler &HCOOH & 0.02-0.08   & 10& 70-90&1-6 &$r_{\rm CB}$\\
(Best-fit)&&0.05&  & 90&5&&0.82\\
\hline
\end{tabular}}
\begin{flushleft}
\tablecomments{In the FERIA code, heights of the envelope and disk are set to be 20 au. The fittings are applied for the cube data of \meta\ and HCOOH with the intensities higher than 3$\sigma$ noise level; 1.5 \mjybeam\ and 1.8 \mjybeam, respectively.
}
\end{flushleft}
\end{table}

\begin{figure}[h!]
\centering
\includegraphics[scale=0.8]{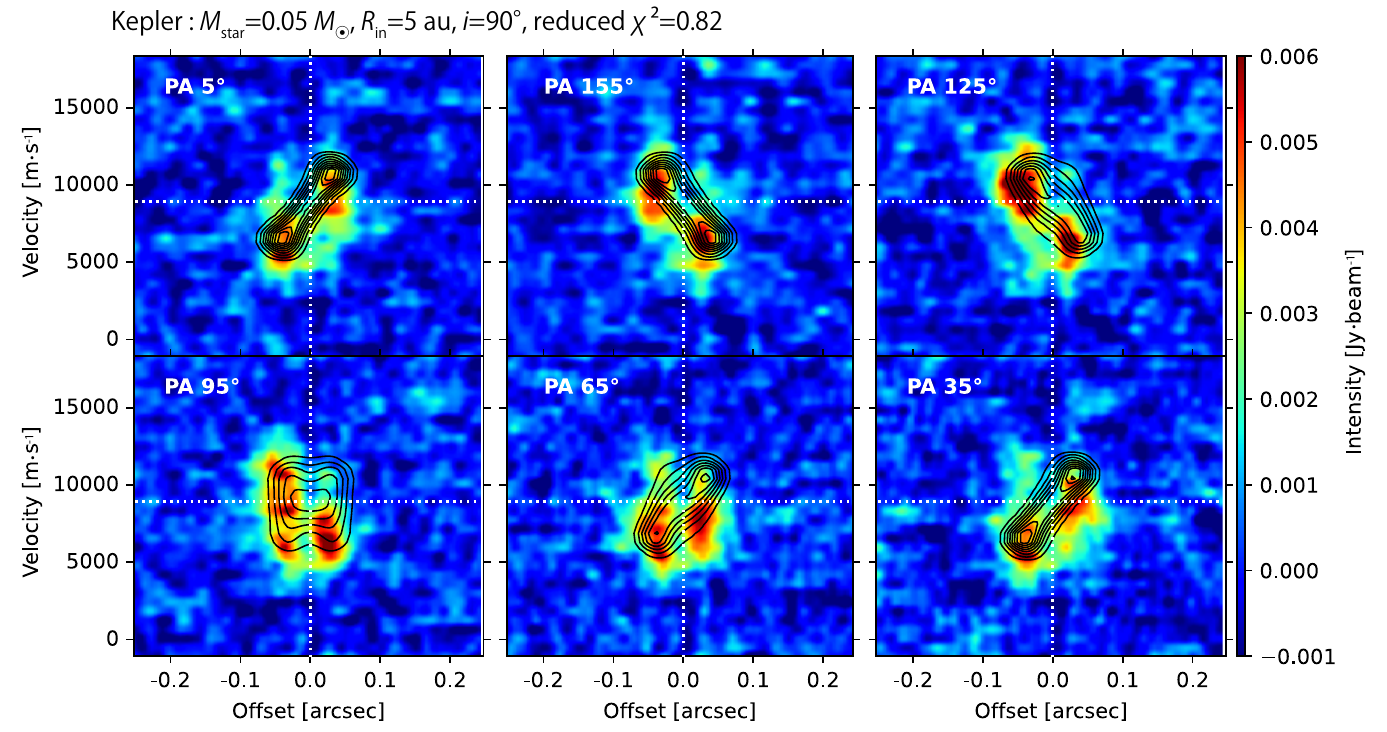}
\caption{PV diagrams of HCOOH (12$_{0,12}-$11$_{0,11}$) in colors and the best-fit model obtained by the $\chi^2$ analysis for the Keplerian disk model with FERIA in contours. The parameters for the model are the protostellar mass of 0.05\sm, the inner radius of 5 au, the outer radius of 10 au, and the inclination of 90\degr\ (90\degr\ for edge-on), where the reduced $\chi^2$ value is 0.82.
Dotted horizontal lines show the systemic velocity of 8.34 \kms \citep{Yen et al.(2015b)}, and dotted vertical lines show the protostar position.
\label{pvhcoohkep}}
\end{figure}

\clearpage
\appendix
\section{Equations for the LTE Calculation}\label{appendix_a}
The column densities of \meta, \metd, \medd, and \medo\ in the envelope direction are derived under the LTE assumption from the observed intensities and the velocity widths (Table \ref{gauss_fit}).
The following formulation is taken from \cite{Okoda et al.(2022)}. Since the dust emission is bright in B335, we explicitly consider the effect of the optical depth of the dust emission ($\tau_{\rm dust}$) as well as that of the line emission ($\tau_{\rm line}$).
For simplicity, we assume the condition that gas and dust are well mixed and the gas temperature is equal to the dust temperature.
In this case, the observed brightness temperature ($T_{\rm obs}$) is represented as follows:

\begin{equation}
T_{\rm obs}=\frac{c^2}{2\nu^2k_{\rm B}}\ \biggl[B_{\nu}(T)+\ {\rm exp\ }\biggl\{-(\tau_{\rm line}+\tau_{\rm dust})\biggr\}\ \biggl\{B_{\nu}(T_{\rm cb})-B_{\nu}(T)\biggr\}-I_{\rm dust}\biggr],
\end{equation}
where $B_{\nu}(T)$ and $B_{\nu}(T_{\rm cb})$ are the Planck function for the source temperature at $T$ and the cosmic microwave background temperature $T_{\rm cb}$, respectively, and $I_{\rm dust}$ is the intensity of the dust continuum emission.
$\tau_{\rm line}$ represents the optical depth of the molecular line, which can be written under the assumption of LTE as:
\begin{equation}\label{line_eq}
\tau_{\rm line}=\frac{8\pi^3 S\mu^2}{3h\Delta v U(T)}\ \biggl\{{\rm exp\ }\biggl(\frac{h\nu}{k_{\rm B}T}\biggr)-1\biggr\}{\ \rm exp\ }\biggl(-\frac{E_u}{k_{\rm B}T}\biggr)\ N,
\end{equation}
where $S$ is the line strength, $\mu$ the dipole moment responsible for the transition, $h$ the Planck constant, $\Delta v$ the full width at half maximum, $U(T)$ the partition function of the molecule at the source temperature $T$. $\nu$ the frequency, $E_u$ the upper-state energy, and $N$ the column density.
The $S\mu^2$ values depend on the transitions (Table \ref{observation}). We employ the experimental values (\sumire) for the lines of \metd\ and \medd\ in Section \ref{column_section}.
On the other hand, $\tau_{\rm dust}$ is given as:
\begin{equation}\label{dust_eq}
\tau_{\rm dust}\sim-\ln\biggl\{\frac{B_{\nu}(T)-I_{\rm dust}}{B_{\nu}(T)}\biggr\},
\end{equation}
assuming that $B_{\nu}(T)\gg B_{\nu}(T_{\rm cb})$. 
The derived $\tau_{\rm dust}$ values are summarized in Table \ref{taudust}.
Further details are presented by \cite{Okoda et al.(2022)}.

\newpage
\setcounter{figure}{0}   
\setcounter{table}{0}   
\section{Gaussian fitting}\label{appendix_b}
\par
\begin{figure*}[h!]
\centering
\includegraphics[scale=0.18]{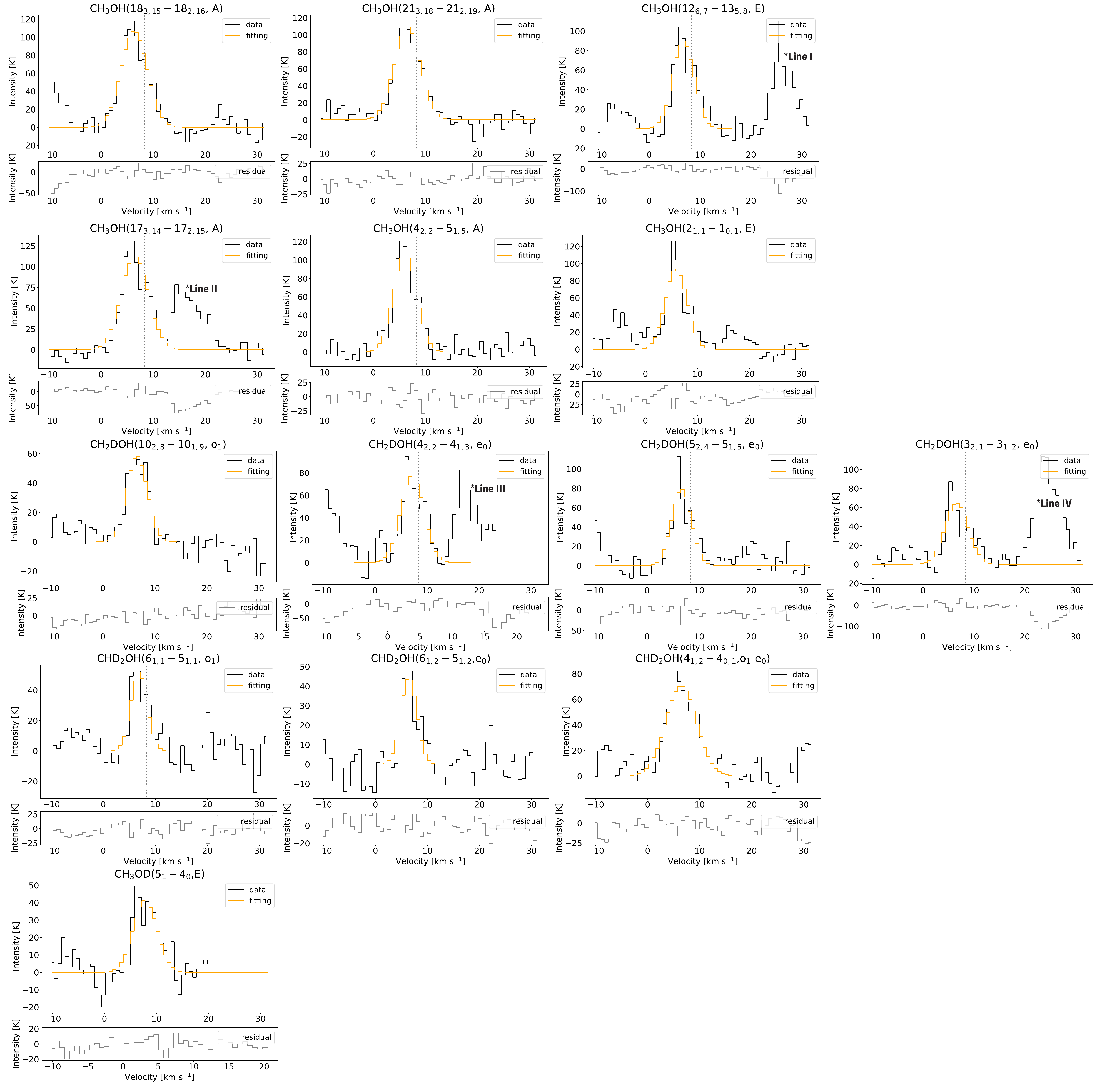}
\caption{Examples of the spectra used for derivation of the column density and the temperature.
The spectra are obtained within the rectangle in Figure \ref{moment0} at the offset of -0\farcs03.
Gaussian fitting spectrum in orange solid line are overlaid with the observed spectrum in black.
The systemic velocity is 8.34 \kms \citep{Yen et al.(2015b)}, as shown in gray dotted lines.
*Line I, II, III and IV are disturbed lines: \metd\ (5$_{2,4}-$5$_{1,5}$, e$_0$), t-HCOOH (11$_{3,8}-$10$_{3,7}$),  CH$_3$CHO (13$_{0,13}-$12$_{0,12}$, A), and \meta\ (18$_{3,15}-$18$_{2,16}$, A), respectively. }
\label{gauss_fit_y-0.03}
\end{figure*}

 Using the method described in Appendix \ref{appendix_a}, we perform the analysis on the observed
intensities for each molecules at each position.
The observed intensities and the velocity widths are obtained by Gaussian fitting.
Figure \ref{gauss_fit_y-0.03} shows examples of the fitting results at the offset of $-$0\farcs03, whose values are summarized in Table \ref{gauss_fit}.
In Figure \ref{gauss_fit_y-0.03}, orange lines and black lines represent the fitting and the observation, respectively.

\counterwithin{table}{section}
\begin{table}[ht]
\centering
\caption{Examples of the Results of Gaussian Best-fit Model at the Offset of $-$0\farcs03} \label{gauss_fit}
\scalebox{1}{
\begin{tabular}{cccccc}
\hline \hline
Line  &  Transition  &  Frequency (GHz)  &  $\delta v$ (\kms)$^a$  &  $I_{\rm peak}$ (K)$^b$  &  $V_{\rm sys}$ (\kms)$^c$  \\
\hline
CH$_3$OH  &  18$_{3,15}-$18$_{2,16}$, A  & 247.610918 & 5.92 $\pm$ 0.33 & 106.99 $\pm$ 5.1 & 6.36 $\pm$ 0.14 \\
CH$_3$OH  &  21$_{3,18}-$21$_{2,19}$, A  & 245.223019 & 5.86 $\pm$ 0.23 & 109.73 $\pm$ 3.76 & 6.54 $\pm$ 0.10 \\
CH$_3$OH  &  12$_{6,7}-$13$_{5,8}$, E   & 261.704409 & 4.96 $\pm$ 0.45 & 89.71 $\pm$ 7.05 & 6.74 $\pm$ 0.19 \\
CH$_3$OH  &  17$_{3,14}-$17$_{2,15}$, A  & 248.282424 & 6.10 $\pm$ 0.76 & 112.8 $\pm$ 12.04 & 6.46 $\pm$ 0.32 \\
CH$_3$OH  &  4$_{2,2}-$5$_{1,5}$, A  & 247.228587 & 5.20 $\pm$ 0.41 & 107.49 $\pm$ 7.37 & 6.10 $\pm$ 0.17 \\
CH$_3$OH  &  2$_{1,1}-$1$_{0,1}$, E   & 261.805675 & 4.84 $\pm$ 0.54 & 94.14 $\pm$ 9.15 & 5.92 $\pm$ 0.23 \\
CH$_2$DOH  &  10$_{2,8}-$10$_{1,9}$, o$_1$  & 244.9888456 & 4.59 $\pm$ 0.29 & 58.33 $\pm$ 3.16 & 6.57 $\pm$ 0.12 \\
CH$_2$DOH  &  4$_{2,2}-$4$_{1,3}$, e$_0$  & 244.8411349 & 5.27 $\pm$ 0.6 & 77.77 $\pm$ 7.64 & 7.18 $\pm$ 0.25 \\
CH$_2$DOH  &  5$_{2,4}-$5$_{1,5}$, e$_0$  & 261.6873662 & 4.45 $\pm$ 0.48 & 79.01 $\pm$ 7.38 & 6.63 $\pm$ 0.20 \\
CH$_2$DOH  &  3$_{2,1}-$3$_{1,2}$, e$_0$  & 247.6257463 & 5.38 $\pm$ 0.76 & 64.61 $\pm$ 7.89 & 6.61 $\pm$ 0.32 \\
CHD$_2$OH  &   6$_{1,1}-$5$_{1,1}$, o$_1$  & 246.1432950 & 3.66 $\pm$ 0.52 & 52.04 $\pm$ 6.44 & 6.87 $\pm$ 0.22 \\
CHD$_2$OH  &  6$_{1,2}-$5$_{1,2}$, e$_0$  & 246.2530390 & 3.59 $\pm$ 0.47 & 44.48 $\pm$ 5.07 & 6.43 $\pm$ 0.20 \\
CHD$_2$OH  &  4$_{1,2}-$4$_{0,1}$, o$_1$-e$_0$   & 247.2524160 & 7.01 $\pm$ 0.45 & 70.58 $\pm$ 3.82 & 6.48 $\pm$ 0.19 \\
CH$_3$OD &  5$_1-$4$_0$, E  & 245.142988 & 5.38 $\pm$ 0.69 & 41.6 $\pm$ 4.63 & 7.84 $\pm$ 0.29 \\
\hline
\end{tabular}}
\begin{flushleft}
\tablecomments{$^a$ Line width of the spectrum. $^b$ Peak intensity of the spectrum. $^c$ Velocity at the peak intensity of the spectrum.}
\end{flushleft}
\end{table}

\counterwithin{figure}{section}

\setcounter{figure}{0}   
\setcounter{table}{0}   
\clearpage
\section{Dust Optical depth}\label{appendix_c}
\par Here, we summarize the dust optical depths derived in the analysis in Table \ref{taudust}. They are evaluated by using Eq. \ref{dust_eq}.

\counterwithin{table}{section}
\begin{table}[ht]
\centering
\caption{Optical Depth and Brightness Temperature of Dust} \label{taudust}
\scalebox{1}{
\begin{tabular}{cccccc}
\hline \hline
 Position& \meta & \metd & \medd & \medo & $T_{\rm dust}$ (K)$^{a}$\\
 \hline
0.1 & 0.04$^{+0.03}_{-0.02}$ & 0.07$^{+0.04}_{-0.05}$ & 0.04$^{+0.03}_{-0.02}$ & 0.04$^{+0.03}_{-0.02}$ & 4.2 \\
0.06 & 0.12$^{+0.02}_{-0.02}$ & 0.15$^{+0.02}_{-0.04}$ & 0.12$^{+0.03}_{-0.02}$ & 0.12$^{+0.03}_{-0.02}$ & 15.4 \\
0.03 & 0.63$^{+0.05}_{-0.04}$ & 0.78$^{+0.04}_{-0.05}$ & 0.63$^{+0.10}_{-0.08}$ & 0.63$^{+0.10}_{-0.08}$ & 74.0 \\
0 & 1.01$^{+0.10}_{-0.03}$ & 1.4$^{+0.1}_{-0.1}$ & 1.0$^{+0.2}_{-0.1}$ & 1.0$^{+0.2}_{-0.1}$ & 136.9 \\
-0.03 & 0.52$^{+0.03}_{-0.03}$ & 0.64$^{+0.05}_{-0.08}$ & 0.52$^{+0.07}_{-0.06}$ & 0.52$^{+0.07}_{-0.06}$ & 73.0 \\
-0.06 & 0.11$^{+0.02}_{-0.02}$ & 0.12$^{+0.03}_{-0.03}$ & 0.11$^{+0.03}_{-0.02}$ & 0.11$^{+0.03}_{-0.02}$ & 14.7 \\
-0.1 & 0.04$^{+0.02}_{-0.03}$ & 0.04$^{+0.03}_{-0.03}$ & 0.04$^{+0.02}_{-0.03}$ & 0.04$^{+0.02}_{-0.03}$ & 4.2 \\
\hline
\end{tabular}}
\begin{flushleft}
\tablecomments{$^{a}$ Averaged brightness temperature of dust continuum emission within each rectangle area shown in Figure \ref{moment0}(a). The noise level of $T_{\rm dust}$ (K) is 2.6 K for a rectangle area (0\farcs03$\times$0\farcs05).}
\end{flushleft}
\end{table}

\setcounter{figure}{0}   
\setcounter{table}{0}   
\clearpage
\section{Optical depths of \meta\ Lines}\label{appendix_d}
\par Here, we summarize the optical depths of the \meta\ lines derived in the analysis in Table \ref{taumeta}. They are evaluated by using Eq. \ref{line_eq}.

\counterwithin{table}{section}
\begin{table}[ht]
\centering
\caption{Optical Depths of the \meta\ Lines} \label{taumeta}
\scalebox{1}{
\begin{tabular}{cccccccc}
\hline \hline
 Position  & 0\farcs1 & 0\farcs06 & 0\farcs03 & 0$''$ & -0\farcs03 & -0\farcs06 & -0\farcs1 \\
 \hline
18$_{3,15}-$18$_{2,16}$, A & 12.646 & 11.566 &  12.161 & 6.725 & 7.382 & 7.898 & 8.963 \\
21$_{3,18}-$21$_{2,19}$, A & 6.841 & 6.504 & 6.814 & 3.097 & 4.415 & 5.099 &  5.005 \\
12$_{6,7}-$13$_{5,8}$, E& 0.783 & 1.899 & 2.834 &  1.739 & 1.924 & 1.734 & 1.061 \\
17$_{3,14}-$17$_{2,15}$, A & 3.932 & 8.621 & 12.446 & 6.294 & 8.898 & 6.938 & 5.165 \\
4$_{2,2}-$5$_{1,5}$, A & 2.942 & 7.341 & 10.819 &  5.207 & 7.767 & 6.176 & 3.934 \\
2$_{1,1}-$1$_{0,1}$, E & 1.162 & 3.121 & 5.213 & 3.377 & 4.360 & 2.895 & 1.576 \\
\hline
\end{tabular}}
\begin{flushleft}
\end{flushleft}
\end{table}

\setcounter{figure}{0}   
\setcounter{table}{0}   
\clearpage
\section{D/H ratios in B335 and other sources}\label{comparison_app}
\par The D/H ratios discussed in Section \ref{medd_metd} are summarized here. We select the 7 low-mass protostellar sources and the low-mass prestellar sources (H-MM1 and L694-2) as samples for comparison. The values of B335 are derived in this paper.

\counterwithin{table}{section}
\begin{table}[ht]
\centering
\caption{Abundances}
\scalebox{1}{
\begin{tabular}{cccc}
\hline \hline
Source & $\lbrack$\medd$\rbrack$/$\lbrack$\metd$\rbrack$ & $\lbrack$\metd$\rbrack$/$\lbrack$\meta$\rbrack$/3 & References$^c$ \\
\hline
B335 & 0.14-0.29$^b$ & 0.03-0.13$^b$ & this work \\
IRAS 16293$-$2422 A & 0.20$\pm$0.07 & 0.028$\pm$0.012 & 1 \\
IRAS 16293$-$2422 B & 0.25$\pm$0.09 & 0.024$\pm$0.009 & 1 \\
B1-c & 0.13$\pm$0.02 & 0.028$\pm$0.09 & 2 \\
Serpens S68N & 0.12$\pm$0.05 & 0.014$\pm$0.006 & 2 \\
NGC1333 IRAS 2A & 0.70$\pm$0.26 & 0.019$\pm$0.01 & 3 \\
NGC1333 IRAS 4A & 0.56$\pm$0.22 & 0.014$\pm$0.008 & 3 \\
HOPS373SW & 0.75 & 0.131 & 4 \\
H-MM1$^a$ & 0.8$\pm$0.4 & 0.06$\pm$0.02 & 5 \\
L694-2$^a$ & 0.5$\pm$0.3 & 0.03$\pm$0.02 & 5 \\
\hline
\end{tabular}}
\begin{flushleft}
\tablecomments{
$^a$ Low-mass prestellar core. $^b$ These values are derived with the experimental value (\sumire). $^c$
1: \cite{Drozdovskaya et al.(2022)}.
2: \cite{van Gelder et al.(2022)}.
3: \cite{Taquet et al.(2019)}.
4: \cite{Lee et al.(2023)}.
5: \cite{Lin et al.(2023)}.}
\end{flushleft}
\end{table}

\end{document}